\documentclass[amsmath,comment,amssymb,twocolumn,showpacs,prb]{revtex4}
\usepackage{graphicx}
\usepackage{amssymb}
\usepackage{epsfig}
\usepackage{dcolumn}
\usepackage{bm}

\newcommand{\be}{\begin{equation}}
\newcommand{\ee}{\end{equation}}
\newcommand{\bea}{\begin{eqnarray}}
\newcommand{\eea}{\end{eqnarray}}

\newcommand{\ra}{\rightarrow}
\renewcommand{\d}{{\rm d }}

\newcommand{\p}{\partial}
\newcommand{\nn}{\nonumber }

\begin{document}

\date{\today}
\title{Mean-field theory for the three-dimensional Coulomb glass}
\author{M.\ M\"uller}
\affiliation{Department of Physics, Rutgers University, Piscataway, New Jersey 08854}
\affiliation{Department of Physics, Harvard University, Cambridge, Massachusetts 02138}
\author{S. Pankov}
\affiliation{National High Magnetic Field Laboratory, Florida State University, Tallahassee, FL 32306}
\pacs{71.23.Cq, 64.70.Pf, 75.10.Nr}

\begin{abstract}
We study the low temperature phase of the 3D Coulomb glass within a
mean field approach which reduces the full
problem to an effective single site model with a non-trivial replica structure. We predict a finite glass transition temperature $T_c$, and a glassy low temperature phase characterized by permanent criticality. The latter is shown to assure the saturation of the Efros-Shklovskii Coulomb gap in the density of states. We find this pseudogap to be universal due to a fixed point in Parisi's flow equations. The latter is given a physical interpretation in terms of a dynamical self-similarity of the system in the long time limit, shedding new light on the concept of effective temperature.
From the low temperature solution we infer properties of the hierarchical energy landscape, which we use to make predictions about the master function governing the aging in relaxation experiments.
 \end{abstract}

\maketitle

\section{Introduction}
\label{s:intro}
In Anderson insulators the Coulomb interactions between localized electrons remain essentially unscreened and give rise to a strongly correlated low temperature state, the so-called electron or Coulomb glass. This state of matter is expected to occur in strongly doped, but insulating semiconductors, in granular metals, as well as in dirty thin metal films. Such systems were long ago predicted to exhibit glassy properties~\cite{Davieslee82,Davieslee84,Gruenewald82,Pollak84} due to their inability to reach the ground state on experimental timescales. The latter leads to experimentally observable out-of-equilibrium phenomena such as the slow relaxation~\cite{benchorin93,Martinez98,Grenet03} of conductivity and compressibility~\cite{DonMonroe87,Monroe90,CugliandoloGiamarchi06}, aging~\cite{vaknin00, Grenet04,Ovadyahu06} as well as memory effects~\cite{Martinez98,vaknin02,LebanonMueller05}. The change of current noise characteristics across the metal insulator transition has also been ascribed to glassy behavior~\cite{Bielec01,Popovic02a}. Further, a series of experiments in doped samples close to the metal insulator transition have shown a variety of glassy features~\cite{Popovic02b,Popovic04,Popovic06,Popovic06b}. 

In recent years there has been growing experimental evidence~\cite{vaknin98} that the glassy behavior observed in some of the above disordered electronic systems is an intrinsic property of the interacting electrons themselves, suggesting that those systems are indeed realizations of Coulomb glasses. In addition there is ample numerical evidence for glassy behavior in such systems~\cite{perez-garrido99,menashe00,Tsigankov03,Grempel04,Kolton05,MuellerLebanon06}.

From a different point of view, insulators with strong Coulomb interactions have long been known to exhibit a prominent suppression in the density of states around the chemical potential, as first argued by Pollak~\cite{Pollak70} and Srinivasan~\cite{Srinivasan71}. Later, Efros and Shklovskii~\cite{efrosshklovskii75} have shown that  a pseudogap $\rho(\epsilon )\leq C \epsilon^{D-1}$ in the local density of states is a necessary prerequisite for any configuration of a $D$-dimensional Coulomb glass to be stable with respect to single particle hops. The presence of this pseudogap leads to a substantial increase of the resistivity from Mott's variable range hopping $R(T)\approx R_0 \exp[(T_M/T)^{1/(D+1)}]$ (assuming a constant density of states) to
Efros-Shklovskii's law, $R(T)\approx R_0 \exp[(T_{\rm ES}/T)^{1/2}]$. Furthermore, it significantly suppresses tunneling at low voltages, as was demonstrated experimentally~\cite{massey95}.

For a long time the relation between these two aspects of strongly interacting Anderson insulators has remained unclear. However, a close analogy with spin glasses suggested that there may be a deeper connection between them~\cite{Anderson78,Davieslee82,pastor99}. Indeed,
Efros' and Shklovskii's (ES) stability argument for the Coulomb gap can also be applied to long range spin glasses, in particular to the Sherrington-Kirkpatrick (SK) model. In that case the argument leads to the conclusion that "the density of states", or more appropriately, the distribution of local fields acting on the spins, must at least have a pseudogap $\rho(\epsilon)\leq |\epsilon|$ to ensure stability with respect to the simultaneous flip of two spins. On the other hand, it is known that for the SK model the presence of a linear pseudogap at low temperature is related to the occurrence of a finite temperature glass transition and the ensuing ergodicity breaking~\cite{BrayMoore79,SommersDupont84}. These observations hold true also for a fully connected electronic model (a fermionic SK model with small quantum fluctuations)~\cite{pastor99} as pointed out by Dobrosavljevic and Pastor. These authors further suggested that a similar relationship might hold between Coulomb gap and the "Coulomb glass" phase in systems with $1/r$ interactions. M{\"u}ller and Ioffe~\cite{MuellerIoffe04} have recently shown that this is indeed true, as will be analyzed more thoroughly in the present paper.

The ES stability argument with respect to single particle hops or spin flips only provides an upper bound for the density of states or the distribution of local fields, while it is difficult to argue rigorously for the actual  saturation of this minimally required power-law suppression. The same is a priori true for long ranged spin glasses.
However, in the case of the SK model, one can actually prove the saturation of the linear bound. Bray and Moore found that there are massless modes in the excitation spectrum around typical minima in phase space~\cite{BrayMoore79}, and that this marginal stability almost implies the presence of a linear pseudogap in $\rho(\epsilon)$. The latter was finally directly shown to be a property of the exact low temperature solution of the SK model~\cite{SommersDupont84,Pankov06}.

For the case of electron glasses, a recently introduced replica mean field approach~\cite{MuellerIoffe04,pankov05} can be used to describe Coulomb correlations in a self-consistent local approximation. This approach not only predicts a finite temperature glass transition, but similarly as for the SK model, an analysis of the low temperature solution shows that the marginality of the glass phase guarantees the presence of a {\em saturated} Efros-Shklovskii Coulomb gap $\rho(E)\propto E^{D-1}$ at low energies~\cite{MuellerIoffe04}.

In this paper, we present a detailed study of this mean field theory for 3D electron glasses and discuss its predictions for experiments and numerical studies. We will frequently compare the formalism for the electron glass to analogous concepts established in spin glasses, and in particular for the SK model. After the introduction of the model and a review
of the mean field approximation in Sec.~\ref{s:singlesite}, we discuss the high temperature phase in Sec.~\ref{s:highT}, and introduce the notion of thermodynamic and instantaneous local densities of states which will be at the focus of the low temperature study.

In Sec.~\ref{s:Glasstransition}, we show that a glass transition occurs due to strong fluctuations in the Thomas-Fermi screening which destroy the exponential screening present in the high temperature phase. This effect leads to a critical (marginally stable) and correlated glassy state throughout the whole low temperature phase.
In this phase, the replica symmetry is spontaneously broken, signaling the presence of many metastable states. The near degeneracy of the latter is at the origin of the marginality whose physical implications will be discussed.

In Sec.~\ref{s:Glassphase} we study the glass phase in detail. We give a physical interpretation of the formalism required to solve Parisi's replica symmetry breaking scheme in the low temperature phase, and prove the permanent marginality of the glass. In Sec.~\ref{s:lowTanalysis} we obtain the local density of states as a function of temperature and disorder from the self-consistent mean field solution. In particular we will see how the phase transition is related to the opening of the Coulomb gap below $T_c$. Sec.~\ref{s:fixedpoint}
focuses on the asymptotic low temperature regime and establishes that the local density of states exhibits the Efros-Shklovskii pseudogap with a saturated exponent $D-1$.
Further, we discuss how lattice effects modify this asymptotic low energy result at intermediate energies to make the gap exponent appear larger ($\approx 2.34$ instead of 2 in 3D).

In Sec.~\ref{s:FixedpointInterpretation} we review the generalization of the fluctuation-dissipation theorem and the occurrence of an effective temperature, and show how it is encoded in Parisi's ultrametric Ansatz and the replica formalism. From the presence of an asymptotic fixed point in the Parisi's renormalization group-like flow equations we obtain a new local interpretation of the effective temperature, and infer a self-similar structure of the long time dynamics. The latter is found to be accompanied by a gradual decrease of the average screening length, $r_{\rm sc}(t)\sim e^2/\kappa T_{\rm eff}(t)$.

As we show in Sec.~\ref{s:Aging}, the structure of the low temperature solution yields non-trivial
information about the structure of phase space in the vicinity of a given metastable state. Combining it with a trap model for dynamics we discuss the possible relevance to aging experiments.
Further possible manifestations of the glassy phase in experiments are addressed in the Discussion (Sec~\ref{s:Discussion}). We discuss limitations of mean field theory and compare the 3D results with numerical approaches, suggesting various future studies to test our predictions. The main results are summarized in the Conclusion (Sec.~\ref{s:conclusion}).

\section{Single site model for the Coulomb glass}
\label{s:singlesite}

\subsection{Disorder average of the lattice Coulomb glass}
We consider the classical lattice model for Coulomb glasses,
\begin{equation}
\beta H=\frac{\beta}{2}\sum_{i\neq j}(n_{i}-K) \frac{e^2}{\kappa r_{ij}}(n_{j}-K)+\beta\sum_{i}n_{i}\epsilon_{i},  \label{Hamiltonian}
\end{equation}
where $n_i \in \{0,1\}$ describes the occupation of site $i$, $K$ is the average site-occupation, $\epsilon_i$ is a random on-site potential and $\kappa$ is the host dielectric constant.
 We restrict our analysis to the case of a half-filled cubic lattice, $K=1/2$, and a Gaussian distribution of the disorder potential. Both assumptions are not crucial to our results, but simplify the further analysis. As long as disorder is much stronger than the interaction between nearest neighbors, the restriction to a cubic lattice should also not be important since the effective low-energy sites will be randomly placed. However, as we will see in Sec.~\ref{s:fixedpoint} lattice effects persist down to relatively low energies, and thus very large disorder is necessary for the glass transition to become insensitive to the underlying lattice.

For the following, it will be convenient to use the Ising spin notation $s_i=2n_i-1 \in \{\pm1\}$.
We consider a cubic lattice with spacing $\ell=1$ which fixes the unit of length.
Replicating the system $n$ times, and averaging over the disorder, we obtain the replica Hamiltonian
\begin{equation}
\beta H_{n}=\frac{1}{2}\sum_{a=1}^n \sum_{i\neq j}s^a_{i}\beta \mathcal{J}_{ij}s^a_{j}-\beta^2 W^2\sum_{a,b=1}^{n}\sum_{i}s^a_{i}\mathcal{I}_{ab}s^b_{i},  \label{repHamiltonian}
\end{equation}
with
\begin{equation}
\mathcal{J}_{ij}=\frac{1}{r_{ij}}, \quad\quad W=\overline{(\epsilon/2)^2}^{1/2},
\label{Jij}
\end{equation}
where the overbar indicates the disorder average.
Here we have chosen $e^2/4\kappa \ell \equiv 1$ as the unit of energy.
${\bf \mathcal{I}}$ denotes a $n\times n$ matrix with all entries equal to 1. In order to describe a {\em quenched} disorder average, the number of replicas has to be sent to $n\ra 0$ eventually. The replicated spin problem (\ref{repHamiltonian}) is amenable to standard high temperature expansions in the Coulomb interactions, as outlined in App.~\ref{app:maptoSS}.

\subsection{Mapping to a single site problem}
In the limit of strong disorder, one finds that the irreducible vertex insertions of the high temperature expansion of App.~\ref{app:maptoSS} are dominated by site-diagonal terms while off-diagonal contributions are suppressed by higher powers of $1/W$. This suggests to look for a self-consistent mapping to an effective single site model which allows us to resum the family of leading diagrams with local vertex insertions.~\cite{MuellerIoffe04,pankov05}

From a physical point of view large disorder suppresses the Thomas Fermi screening by quenching the electrons. This preserves the long range of the Coulomb interactions and hence a large number of effective neighbors for a given site. This situation is favorable for a cavity or mean-field approach where the behavior of a single site and its environment are described in a self-consistent manner.

On a technical level, a similar approximation to the above "locator" approximation is made in dynamical mean field theories~\cite{DMFT}, which is usually justified by invoking the limit of large dimensionality. In the present problem, however, we can use the long range nature of the essentially unscreened Coulomb interaction to justify the mean field approximation. Since we are treating a classical, thermodynamic rather than a quantum dynamic problem, we do not have a frequency dependence of the replica diagonal part of the self-energy. Instead, we obtain a non-trivial structure in the replica sector. It is well-known that the latter encodes information about dynamic behavior in the asymptotic long time limit, assuming a generic Langevin dynamics of the spins~\cite{fn1}.

\subsection{Self-consistent description}
We seek to map the full lattice problem onto a self-consistent single site model with the effective Hamiltonian~\cite{fn2}
\begin{equation}
\beta H_{0}(\{s_{\alpha }\})=-\frac{\beta^2}{2}\sum_{a,b}s_{a}(\Lambda_{ab} + W^2\mathcal{I}_{ab})s_{b},  \label{H0}
\end{equation}
whose exact solution is supposed to resum all local (cactus-like) diagrams of the original lattice problem. In order to make the mapping self-consistent we require that both the local irreducible polarizability $\Pi_{ab}$ (i.e., the irreducible two-leg vertex of the diagrammatic expansion, cf.~Eq.~(\ref{M2Pi})), and the local two-point functions $\langle s_{i,a}s_{i,b}\rangle$ (i.e., the overlap $Q_{ab}$ in spin terminology) be the same in the two models. As derived in detail in App.~\ref{app:maptoSS} this leads to the self-consistency conditions
\begin{eqnarray}
&&Q_{ab}=\langle s_{a}s_{b}\rangle_{H_0} =
\left(-\beta^2 {\bf \Lambda}-\beta/{\bf \Pi}\right)^{-1}_{ab}
\label{SP} \\
&&=\frac{1}{N} \sum_{i}\langle s_{i,a}s_{i,b}\rangle_{H_n} =\int \frac{d^3k}{(2\pi)^3} \left(\beta \mathcal{J}_k -\beta/{\bf \Pi}\right)^{-1}_{ab}.
\nonumber
\end{eqnarray}%
where brackets denote a thermal average and $\mathcal{J}_k$ is the Fourier transform of the Coulomb interaction (\ref{Jij}), $\mathcal{J}_k=4\pi /k^2$ for $k\rightarrow 0$. Eq.~(\ref{SP}) holds for all $(a,b)$, with the obvious constraint for the diagonal elements $Q_{aa}\equiv \tilde{Q}=1$.

Using the first line in (\ref{SP}) to substitute $\beta/{\bf \Pi}$ by ${\bf Q}^{-1}+ \beta^2 {\bf \Lambda}$, 
we obtain the self-consistency equation relating overlap ${\bf Q}$ and effective coupling ${\bf \Lambda}$,
\begin{eqnarray}
\label{SC1}
Q_{ab}=\langle s_{a}s_{b}\rangle_{H_0} =\int \frac{d^3k}{(2\pi)^3} \left(\beta \mathcal{J}_k+{\bf Q}^{-1}+\beta^2 {\bf \Lambda} \right)^{-1}_{ab}.
\end{eqnarray}%

We rewrite Eq.~(\ref{SC1}) in more compact form,
\begin{eqnarray}
\label{SC2}
\beta {\bf Q}=\chi_1[(\beta {\bf Q})^{-1}+\beta {\bf \Lambda}],
\end{eqnarray}%
introducing the response function
\begin{equation}
\label{chi}
\chi_1(\rho^2)=\int \frac{d^3k}{(2\pi)^3} \frac{1}{\mathcal{J}_k+\rho^2}.
\end{equation}
Notice that the argument $\rho^2$ takes the place of an inverse polarizability operator. In the context of Coulomb interactions it has the interpretation of the square of an average Thomas-Fermi screening length, whence the suggestive notation $\rho$.

For a strongly suppressed polarizability, $\rho \gg 1$, e.g., due to large disorder $W$ or at low temperature, one obtains the asymptotic behavior
\bea
\label{fasym}
\chi_1(\rho^2)&=&\frac{1}{\rho^2}+\int \frac{d^3k}{(2\pi)^3} \frac{1}{\rho^4} \frac{\mathcal{J}_k^2}{\mathcal{J}_k+\rho^2}\nn\\
&\approx& \frac{1}{\rho^2}\left(1+\frac{2\sqrt{\pi}}{\rho^3}+\frac{C_{\rm latt}}{\rho^4}+o\left(\frac{1}{\rho^{4}}\right)\right).
\eea
We have used the fact that the absence of self-interactions ensures ${\cal J}(r=0)=\int_k{\mathcal J}_{k}=0$. On a lattice the integrals over $k$ are cut off on short scales $k\sim 1/\ell$, in such a way as to preserve the normalization
$\int_k \equiv \int d^3k/(2\pi)^3=1$.
Note that the second term in (\ref{fasym}) is dominated by the long range properties at small $k$ of the interactions, ${\cal J}_k\ra 4\pi/k^2$ and is thus universal, i.e., independent of the considered lattice. The higher order terms are sensitive to short length scales, the leading corrections being given by
\bea
\label{Nonuniversalcorrection}
C_{\rm latt}&=&\int_{k\in {\rm BZ}} {\cal J}_k^2 \,\frac{d^3k}{(2\pi)^3}-\int_{k\in \mathbb{R}^3 } \left(\frac{4\pi}{k^2}\right)^2\,\frac{d^3k}{(2\pi)^3}\nn\\
&=&\lim_{k_0\to 0}\int_{k\in {\rm BZ},\,k>k_0} \mathcal{J}_k^2 \,\frac{d^3k}{(2\pi)^3}-\frac{8}{k_0},
\eea
where ${\rm BZ}$ indicates the first Brillouin zone of the considered lattice.
Below we will consider the case of a cubic lattice for which one finds $C_{\rm latt}
=-8.9555$
using the lattice Fourier transform ${\cal J}_k$ of App.~\ref{app:LatticeFT}.

\section{High temperature phase}
\label{s:highT}
At high temperatures, the system is in a unique thermodynamic state (a single ergodic compartment), as reflected  by the unbroken replica symmetry of the overlap and coupling matrices in the self-consistent local approximation ($Q_{a\neq b}=Q_{\rm RS}$, $\Lambda_{a\neq b}=\Lambda_{\rm RS}$).

The physical content of this high temperature solution is rather easy to understand: Due to the random onsite fields, $\epsilon_i$, each site carries an average charge or "magnetization" in spin language, $q_i=\langle s_i\rangle/2 \equiv m_i/2$.
Therefore, even at high temperature, two equilibrated copies $\alpha$ and $\beta$ of the same disordered sample have a non-vanishing average similarity, i.e., a positive overlap $Q^{\alpha\beta}$ of their site occupation pattern,
\bea
Q^{\alpha\beta}=\frac{1}{N}\sum_i\langle s_i^\alpha s_i^\beta\rangle=\frac{1}{N}\sum_i m_i^\alpha m_i^\beta.
\eea
In the single site approximation this translates into a finite replica off-diagonal overlap $Q^{\alpha\beta}=Q_{a\neq b}$.

The Coulomb repulsion from the average charges $q_i$ augments the original Gaussian disorder on other sites and makes the distribution of local fields wider. The self-consistent approximation treats this extra disorder as a Gaussian distribution with width $\Lambda_{\rm RS}$. This physics is contained in Eqs.~(\ref{SC1},\ref{SC2}) which can be rewritten for the replica symmetric case in the form
\bea
\label{RSequations1}
Q_{\rm RS}&=&\langle s_a s_b\rangle_{H_0} =\int\d y \,P_{{\rm RS}}(y; \Lambda_{\rm RS}) \tanh^2(\beta y),\\
\label{RSequations2}
\Lambda_{\rm RS}&=&\psi\left[(r^{\rm RS}_{\rm sc})^2\right]Q_{\rm RS},
\eea
where we have introduced the distribution of effective local fields, $P_{\rm RS}(y)$, and the screening radius, $r^{\rm RS}_{\rm sc}$,
\bea
P_{\rm RS}(y; \Lambda_{\rm RS})&=&\frac{\exp[-y^2/2(W^2+\Lambda_{\rm RS})]}{\sqrt{2 \pi (W^2+\Lambda_{\rm RS})}},
\label{PRS}\\
\left(r^{\rm RS}_{\rm sc}\right)^2&=&\chi_1^{-1}[\beta(1-Q_{\rm RS})].\label{rscRS}
\eea
We have also defined the function
\bea
\label{psi}
\psi(\rho^2)&=&\frac{1}{\chi_1^{2}(\rho^2)}-\frac{1}{\chi_2(\rho^2)},
\eea
where
\begin{equation}
\label{chi2}
\chi_2(\rho^2)\equiv -\chi_1'(\rho^2)=\int \frac{d^3k}{(2\pi)^3} \frac{1}{(\mathcal{J}_k+\rho^2)^2}.
\end{equation}
$\psi$ contains the crucial information about the Coulomb interactions on the original lattice. A physical interpretation of this function will be given further below.

The asymptotics of $\psi(\rho^2)$ for large screening radius $\rho$ is obtained as
\bea
\label{psi2}
\psi(\rho^2)&\approx& \left(\pi \rho^2\right)^{1/2}+2C_{\rm latt}+ o(1).
\eea
The exponent $\mu=1/2$ of the leading term reflects the universal long range tail of the Coulomb interactions in 3D. For general long range repulsive interactions ${\cal J}(r)\propto 1/r^\nu$ in $D$ dimensions it generalizes to $\mu=1-\nu/(D-\nu)$.

Note that the high temperature solution of Eqs.~(\ref{RSequations1},\ref{RSequations2}) determines self-consistently the Gaussian extra disorder $\Lambda_{\rm RS}$.
More insight into the meaning of (\ref{RSequations2}) can be gained  from the analogous equation for the high temperature phase of the Sherrington-Kirkpatrick spin glass in a random field which we outline in App.~\ref{app:SK}. For that model one finds the exact identity $\psi(\rho^2)\equiv 1$, and consequently the overlap and effective coupling matrices coincide, $\Lambda_{\rm RS}=Q_{\rm RS}$. This is physically transparent considering the explicit expression of the variance of the additional random fields created by the site magnetizations $m_j$,
\bea
\Lambda_{\rm RS}=\frac{1}{N}\sum_i\overline{\sum_j(J_{ij}m_j)^2}=\frac{1}{N}\sum_j m_j^2=Q_{\rm RS}.
\eea

The case of the electron glass is more involved because the equivalent expression with bare interactions ${\cal J}_{ij}= 1/r_{ij}$ diverges. Instead, one has to take into account that interactions are screened by other charges, however, without double counting direct interactions and screening contributions. The locator approximation provides a compact way to solve this problem in a self-consistent manner. Indeed, a formal expansion of the self-consistent $\Lambda_{\rm RS}$ in powers of ${\cal J}$ shows that the lowest order term $\Lambda^{(0)}_{\rm RS}=\overline{\sum_{j,k} (J_{ij}m_j J_{ik} m_k)}$ is modified by screening corrections
\bea
\delta\Lambda_{\rm RS}=\overline{\sum_{j,k,l} (J_{ij}m_j)J_{ik} \beta(1-m_k^2) J_{kl}}+\dots
\eea
which have to be resummed to yield a finite result in the self-consistent local approximation.

\subsection{The density of states}
\label{ss:DOShighT}
The above high temperature formalism yields a Gaussian distribution $P_{\rm RS}(y)$ of effective local fields $y$. We refer to them as "thermodynamic" fields in the sense that the average occupation of a site is given by $\langle s_i\rangle =m_i=\tanh(\beta y_i)$. This field distribution controls the compressibility (charge susceptibility)
\bea
\label{susc}
\kappa_C=\beta(1-Q_{RS})=\beta \int \d y \frac{P_{\rm RS}(y)}{\cosh^2(\beta y)}.
\eea
The field $y_i$ describes the energy to flip the occupation on the site $i$,
 including local rearrangements of neighboring particles which tend to lower the energy cost as compared to an isolated flip. Note that in a glassy state such local relaxations differ from a global relaxation for which large portions of the system need to rearrange to best accommodate the flipped site. More precisely, one should think of $y_i$ as taking into account relaxations within the phase space valley corresponding to the current local metastable state.

\subsection{Onsager term}
\label{ss:Onsager}
For certain experiments in strongly insulating materials, such as photoemission or tunneling from a broad junction, what matters is rather the distribution of "instantaneous" local fields $h_i=dH/d n_i=\epsilon_i+\sum_j e^2/\kappa r_{ij}n_j$ (the cost of isolated flips). While multi-particle processes may in principle occur, the cost in quantum action associated with the hopping of electrons other than the one which is emitted or tunnels largely exceeds the corresponding gain in density of states. The contribution of multi-particle processes to such observables is thus greatly suppressed in strongly insulating classical Coulomb glasses.

With some additional reasoning, we can obtain the distribution of "instantaneous" local fields from the above formalism, too.
In spin language, the time-averaged local field $\left\langle h_i \right\rangle=-\sum_{j\neq i}\mathcal{J}_{ij}m_j$, and the "thermodynamic" field $\beta y_i = \tanh^{-1}(m_i)$ differ in general, because the latter includes local relaxation processes in the environment, while the first describes solely the average energy cost to flip the occupation of the single site $i$. Consequently, if we hold the spin $s_0$ fixed at site $0$, the
average local field it sees is larger than the thermodynamic field $y_0$
by the polarization response of the environment,
\begin{equation}  \label{TAPrel}
\left\langle h_0 \right\rangle_{s=s_0} = y_0+ s_0 h_O.
\end{equation}
The term $h_O$ is known as Onsager's back reaction, familiar from
similar equations obtained by Thouless-Anderson-Palmer (TAP) for the SK-model~\cite{ThoulessAnderson77}. In that model, $h_O$ is given by the sum over all local cactus diagrams starting with two legs at site $0$. Those are simply resummed by $h_O=
\sum_{i\neq 0} \mathcal{J}^2_{0i} \Pi_{i}$, with the irreducible polarizability $\Pi_i\approx dm_i/dh_i=\beta(1-m_i^2)$ (up to terms of order $1/N$). Using $\overline{{\cal J}_{0i}^2}=1/N$ and the Edwards Anderson self-overlap $Q_{\rm EA}=N^{-1}\sum _i m_i^2$, one finds $h_O= \beta(1-Q_{\rm EA})$.

In the case of Coulomb glasses, the expression for $h_O$ has to be generalized since the equivalent of the above expression diverges for ${\cal J}_{ij}=1/r_{ij}$. Indeed, we have to sum over all higher
order ring diagrams~\cite{Srinivasan71},
\begin{equation}
\label{generalizedOnsager2}
h_O= \sum_{j_1 \neq 0} \mathcal{J}_{ij_1} \Pi_{j_1}\mathcal{J}_{j_1i}-\sum_{j_1,
j_2 \neq 0} \mathcal{J}_{ij_1} \Pi_{j_1}\mathcal{J}_{j_1j_2}\Pi_{j_2}\mathcal{J}%
_{j_2i} +\dots
\end{equation}
where the $\Pi_j$'s are local (site-dependent) irreducible polarizabilities in the presence of the fixed spin at site $i$. 

In the spirit of the present mean field approach, we approximate $\Pi_j$ by their site average which is justified by the large number of sites contributing to the reaction. To leading order in large disorder and at low temperatures we have $\overline{\Pi_j}\approx 1/r_{\rm sc}^2$. We may then perform the sum over ring diagrams and obtain
\bea
 \label{generalizedOnsager}
h_O= 
{\rm Tr} \frac{\mathcal{J}^2}{\mathcal{J}+r^{2}_{\rm sc}}
\approx \frac{2\sqrt{\pi}}{r_{\rm sc}},
\eea
with the Thomas-Fermi screening radius $r_{\rm sc}$, as introduced in Eq.~(\ref{rscRS}) for the high temperature phase. It will be generalized later to the low temperature regime (cf.,~Eq.~(\ref{rsc})).
The approximation is valid for a strongly suppressed susceptibility, where $r_{\rm sc}$ is large.
In App.~\ref{app:Onsagerterm} we derive the more precise expression $h_O=\beta(\Lambda_{aa}-\Lambda_{b\ra a})=\chi_1^{-1}(\kappa_C)-1/\kappa_C$, taking properly into account the exclusion of site $i$ from the ring diagrams. Note that for the SK model where $\Lambda\equiv Q$, this form reduces to the expression given above.

In the high temperature, replica symmetric phase of a strongly disordered sample the irreducible polarizability is approximately equal to the bare local susceptibility $\Pi\approx \chi_{\rm loc}\approx 2/\sqrt{2 \pi}W$, and thus
\bea
\label{Onsager RS}
h_O^{\rm RS} \approx \frac{2(2 \pi)^{1/4}}{W^{1/2}}.
\eea
When the Onsager back reaction becomes comparable to the temperature, the spins develop strong correlations with their environment. We will see in the next section that these correlations actually induce a glass transition at $T_c \sim h_O^{\rm RS}$.

\subsection{Distribution of instantaneous fields}
\label{ss:Instfields}
With Eq.~(\ref{PRS}) we obtained the distribution of {\em thermodynamic} fields, $P(y)$, from the self-consistent solution of the mean field approach.
In this subsection, we show how to obtain the distribution of local
{\em instantaneous} fields $h_i$, using the insight from the previous subsection. For the joint distribution of $h_i$ and the orientation of the spin $s_i$, we have by definition
\begin{eqnarray}
\label{Pofhands}
P(h,s)&=&\frac{1}{V}\sum_i\int \frac{d\lambda}{2\pi}\left\langle \exp\left[i\lambda
h-i\lambda\sum_j \mathcal{J}_{ij}s_j\right] \delta_{s_i s}\right\rangle  \nonumber
\\
&\approx&\frac{1}{V}\sum_i\int \frac{d\lambda}{2\pi} \exp\left[i\lambda h-i\lambda\sum_j
\left\langle \mathcal{J}_{ij}s_j\right\rangle_{s_i=s}\right] \nn\\
&&\quad \times \exp\left[-\frac{\lambda^2}{2}\sum_{j,k} \mathcal{J}_{ij}\left\langle s_j s_k\right\rangle_{c,s_i=s} \mathcal{J}_{ki}\right].
\end{eqnarray}
We have only retained the first two cumulants, to be consistent within the
locator approximation. From the TAP-like equations (\ref{TAPrel}) we
may identify the first cumulant as
\begin{equation}
\left\langle \mathcal{J}_{ij}s_j\right\rangle_{s_i=s}=\left\langle h_{i}\right\rangle_{s_i=s} = y_i+s h_O.  \label{TAPeq}
\end{equation}%
The second cumulant is almost insensitive to the value of the spin at site $i$, and evaluates to
\begin{eqnarray}
&&\sum_{j,k}\mathcal{J}_{ij}\left\langle s_j s_k\right\rangle_{c,s_i=s} \mathcal{J}%
_{ki}\approx \left[\mathcal{J}\frac{1}{\beta(\mathcal{J}+r^2_{\rm sc})}\mathcal{J}\right]%
_{ii}\\
&&\quad\quad = \frac{1}{\beta}{\rm Tr}\left[\frac{\mathcal{J}^2}{\mathcal{J}+r^2_{\rm sc}}\right]=
 \frac{h_O}{\beta}
\end{eqnarray}
in the locator approximation. A closer analysis of the Onsager term (cf. App.~\ref{app:Onsagerterm}), shows that this relation is essentially exact.

The righthand side of Eq.~(\ref{Pofhands}) can now be expressed in terms of averages over the joint probability distribution for $(y_i,s_i)$, given by $P(y)\exp(s \beta y)/2\cosh(\beta y)$,
\bea
P(h,s)&=&\int dy P(y) \left[\frac{\exp(s \beta y)}{2\cosh(\beta y/2)}\right.\\
&& \left.\times \int \frac{d\lambda}{%
2\pi} e^{i\lambda (h-y-s h_O)}
 e^{-\lambda^2 h_O/2\beta}%
\right]\nn\\
&=&\int dy P(y) \frac{e^{s \beta y}}{2\cosh(\beta y)} \frac{\exp[
-\frac{\beta(h-y-sh_O)^2}{2h_O}]}{\sqrt{2\pi h_O/\beta}}.\nn
\eea
Summing over $s=\pm 1$ we finally obtain  the desired distribution of instantaneous fields,
\begin{equation}
\label{PofhfromPofy}
P(h)=\int dy P(y) \frac{\cosh(\beta h)}{\cosh(\beta y)} \frac{\exp[-\frac{\beta(h-y)^2}{2h_O}-\frac{\beta h_O}{2}]}{\sqrt{2\pi h_O/\beta}}.
\end{equation}
This relation generalizes an expression derived for the SK model in Ref.~\onlinecite{Thomsen86}.
Notice that for fields $h,y\gg h_O, T$ the two distributions essentially differ by a shift, $P(h)=P(y=h-h_O)$.

\subsection{Correlations in the instantaneous field distribution $P(h)$}
In the high temperature phase
the distribution of thermodynamic fields $P(y)$ is a featureless Gaussian, cf., Eq.~(\ref{PRS}). Yet, due to the Onsager back reaction, a correlation hole starts developing in the distribution
of instantaneous fields $P(h)$, well before the transition to a strongly correlated glassy state occurs at $T_c$.
The high temperature correlation hole in $P(h)$ is thus not directly related to the low temperature glass phase, but simply reflects particle-hole correlations
in the liquid Coulomb "plasma". This was discussed in detail in Ref.~\onlinecite{pankov05}, which we follow here in referring to this correlation hole as a "plasma dip".
Note that while at the transition nothing particular happens to $P(h)$, $T_c$ marks the opening of a pseudogap in $P(y)$ which is related to the breaking of ergodicity, as we will see in more detail in the next section.

\begin{figure}
\includegraphics[width=3.0in]{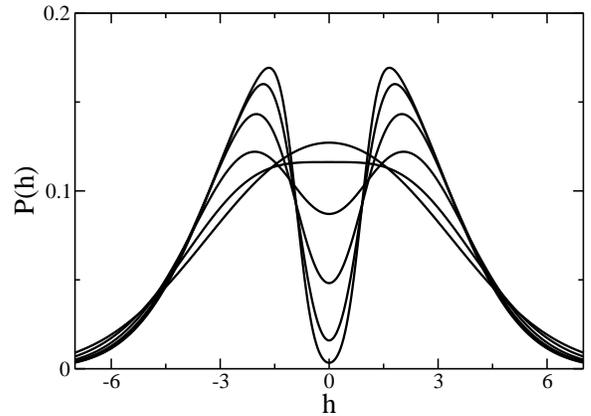}
\caption{Evolution of the correlation hole (plasma dip) in the distribution of the instantaneous fields $P(h)$ for moderate disorder $W=2$. The curves are plotted for the
inverse temperatures $\beta=0.25, 0.5, 1, 2, 4, 7$ (from top to bottom at $h=0$) above the glass transition ($\beta_c = 7.149$). The plasma dip starts opening as a perturbation of order $\mathcal{O}(T_c/T)$ at high temperature and develops into a significant depletion already for $T>T_c$. 
}
\label{f:PhhighT}
\end{figure}

The high temperature distribution $P(h)$ peaks at finite fields of the order of $h_O$ (see Fig.~\ref{f:PhhighT}). The peak at negative fields is due to
sites $i$ occupied by electrons which attract holes to neighboring
sites, inducing in turn a positive potential on the site $i$. This is again a manifestation of Onsager's back reaction. The same reasoning applies to the peak at
positive fields with the role of holes and electrons interchanged.
These correlations are present at all
temperatures, but they are smeared by thermal fluctuations when
 $T\gg h_O^{\rm RS}$.
Note that the apparent parabolic shape of the plasma dip above the glass transition has little to do with the \textit{universal} Efros-Shklovskii gap $\rho(E)\sim E^2$ which we will see emerging in the low temperature glassy phase, cf.~Fig.~\ref{f:Pyandh_w=2}.

The plasma dip persists also in strong disorder. In fact, the replica symmetric mean field theory predicts a universal shape for $P(h)$ in the limit of very strong disorder and temperatures of the order of $T \gtrsim T_c$. This prediction relies on the suppression of screening in the presence of large disorder, and applies to instantaneous fields in the regime $h\sim T_c$, much below the scale of nearest neighbor interactions.
Using Eq.~(\ref{PofhfromPofy}) and anticipating $h_O^{\rm RS}=3 T_c$ (cf. Eq.~(\ref{TcvshO})), valid for large disorder, one finds
\bea
\label{plargewuniv2}
&&\tilde P(\tilde{h})=\\
&&\quad \frac{\gamma }{2\pi\sqrt{3}}\int d\tilde{y}\, \frac{\cosh(\gamma \,\tilde{h})}{\cosh(\gamma \,\tilde{y})}
\exp\left[-\frac{3}{2}-\frac{\gamma ^2}{6}(\tilde{h}-\tilde{y})^2\right].\nn
\eea

\begin{figure}
\includegraphics[width=3.0in]{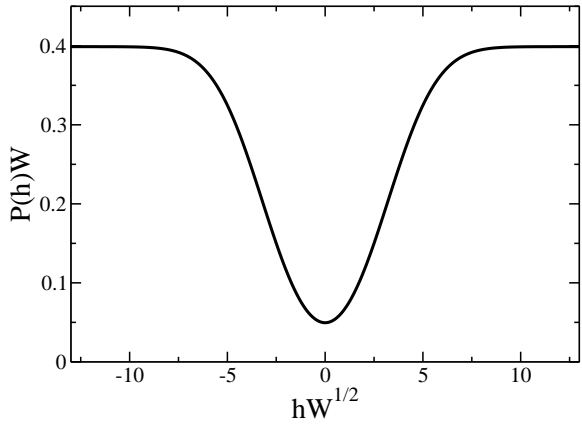}
\caption{Universal scaling of the plasma dip in the distribution of instantaneous fields in large disorder 
at $T=T_c$.}
\label{f:plargew}
\end{figure}

The universal shape of $P(h)$ is illustrated in Fig.~(\ref{f:plargew}) where we plot the scaled function $\tilde P(\tilde{h})\equiv W P(h=\tilde{h}/\sqrt{W})$,
evaluated at the glass transition $1/T_c\equiv \beta_c=\gamma \sqrt{W}$ (with $\gamma$ given in Eq.~(\ref{factorgamma}) below).

\section{Glass transition}
\label{s:Glasstransition}
The high temperature phase of the electron liquid exhibits a glass instability when the temperature becomes comparable to the Onsager term $h_O$.
At large disorder, one finds the scaling $T_c\sim h_O\sim  1/\sqrt{W}$. This is identical to the estimate~\cite{pankov05} of the width of the Efros-Shklovskii Coulomb gap, which follows from equating the supposedly saturated pseudogap $\rho(\epsilon)\sim \epsilon^2$ to the bare density of states~\cite{Pollakeedisorder} $\rho_0\sim 1/W$.
This illustrates the fact that both phenomena are based on the strong electron-electron correlations building up below the scale $T_c$.

An analogous estimate of the critical temperature for the SK-model in strong random fields predicts a transition at $T_c \sim h_O\sim 1/W$, which is indeed confirmed by the exact expression for the Almeida-Thouless instability (see App.~\ref{app:SK}).

\subsection{The glass instability}
The instability at $T_c$ can be understood in various ways. Technically, the simplest way to find the instability is to study the free energy of the effective single site model,
\bea
\label{freeenergy}
n\beta F({\bf \Lambda})&=&\frac{1}{2}\sum_{a}\left[ \ln (-\beta^2{\bf \Lambda}-\beta/{\bf \Pi})\right]_{aa}\\
&&-\frac{1}{2}\sum_a\left[\int_k \ln \left( \beta \mathcal{J}_k-\beta/{\bf \Pi} \right)\right]_{aa}\nn\\
&& \hspace{-0.5cm} -\ln \left[ \sum_{s_{\alpha }}\exp\left(-\frac{\beta^2}{2}%
\sum_{a,b }s_{a} (\Lambda_{ab}+W^{2}\mathcal{I}_{ab})s_b \right)\right], \nn
\eea
where ${\bf \Pi}={\bf \Pi}({\bf \Lambda})$ is determined from the self-consistency Eq.~(\ref{SP}), equivalent to $\partial F/\partial {\bf \Pi}=0$. This free energy expression can be obtained either by "integrating" the saddle point equations~(\ref{SP}), or from a Baym-Kadanoff functional, restricted to diagrams with a purely local self-energy.

One finds that the RS solution becomes unstable to replicon fluctuations around the replica symmetric high temperature solution ($\Lambda_{ab}=\Lambda_{ab}^{\rm RS}+\delta \Lambda_{ab}$ with $\sum_b\delta \Lambda_{ab}=0$ and $\delta \Lambda_{aa}=0$) when
\begin{eqnarray}
\label{RSinstab}
\beta^2 \psi(r_{\rm sc}^{\rm RS})\int_{-\infty }^{\infty }dy \frac{P(y; \Lambda^{\rm RS})}{\cosh^4 \left( \beta y\right)}  =1.
\end{eqnarray}%

In the large disorder limit one obtains
\bea
\label{factorgamma}
\beta_c=\gamma \sqrt{W}, \quad \textrm{with}\quad  \gamma=3/2(2\pi)^{1/4}.
\eea
Upon restoring lattice units this reads
\bea
\label{Tc}
T_c&=& \frac{e^2/\kappa\ell}{6(2/\pi)^{1/4}\left(\frac{2W}{e^2/\kappa\ell}\right)^{1/2}}=\frac{e^2/\kappa \sqrt{\nu_0 e^2/\kappa}}{6(2/\pi)^{1/4}},
\eea
where $\nu_0=(2W\,\ell^3)^{-1}$ is the "bare" density of electronic states $\epsilon_i$. Using Eqs.~(\ref{SC1},\ref{SC2}) it is straightforward to establish that in the limit of large disorder
\bea
\label{TcvshO}
T_c =h_O^{\rm RS}/3\quad \textrm{for} \quad W\to \infty,
\eea
confirming the previous assertion that the glass transition temperature scales like the high temperature Onsager reaction.

\subsection{Critical fluctuations and breakdown of homogeneous screening}
A more physical understanding of the instability can be obtained from an analysis of the connected spin-spin correlation function $\langle s_i s_j\rangle_c$ of the original lattice model before disorder averaging. According to the above arguments, its high temperature series is dominated by chain-like screening diagrams with local polarizability insertions $\Pi_i$,
\bea
\label{averagepropagator}
\beta \langle s_i s_j\rangle_c&&=\beta(\langle s_i s_j\rangle-\langle s_i\rangle\langle s_j\rangle)\\
&&\approx \Pi_i\left[-{\cal J}_{ij}+\sum_{l} {\cal J}_{i l}\Pi_l {\cal J}_{l j}-+\dots\right]\Pi_j.\nn
\eea
In large disorder, we have approximately $\Pi_i\approx  \beta(1-\langle s_i\rangle^2) \approx \beta/\cosh^2(\beta \tilde{\epsilon}_i)$, where $\tilde{\epsilon}_i$ is the local disorder of site $i$, renormalized by the extra disorder due to Hartree interactions with other sites (described by tree-like dressings of the bare vertices). To leading order in large disorder, the renormalized disorder is essentially the same as the thermodynamic field $y_i$, with distribution $P_{\rm RS}$ (\ref{PRS}).

Calculating the disorder average of the geometric series (\ref{averagepropagator}) yields an exponential decrease of the {\em average} correlations, with distance,
\bea
\overline{\langle s(0) s(r)\rangle_c} &=& \beta^{-1}\overline{\Pi}^2\int_k \frac{\exp[ikr]}{{\cal J}_k^{-1} +\overline{\Pi}}\nn\\
&=&\beta^{-1} \overline{\Pi}^2 \frac{e^{-r/R_0}}{r}.
\eea
where $\overline{\Pi}\approx 2/\sqrt{2\pi}W$ denotes the polarizability averaged over the (renormalized) local disorder $\tilde{\epsilon}$. The average screening radius is given by $R_0^2=1/4\pi \overline{\Pi} \sim W$ which grows with disorder. The self-consistent approximation in the high temperature phase resums diagrams renormalizing the local polarizabilities and predicts $R_0^2=(r^{\rm RS}_{\rm sc})^2/4\pi$, which improves on the subleading corrections in $1/W$.

\begin{figure}[tbp]
\resizebox{7.5 cm}{!}{\includegraphics{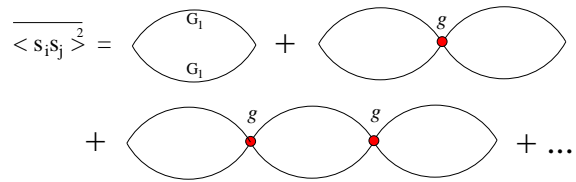}}
\caption{(Color online) Typical ladder diagrams contributing to the variance of the spin-spin correlation function. The single lines represent disorder averaged propagators, the vertices correspond to the variance of the irreducible polarizability.}
\label{f:squaredcorrelator}
\end{figure}

The above result may be misleading, however, since the correlation functions can fluctuate strongly and might not even have a definite sign. In order to capture such effects one should compute the variance of the correlation function. This requires the summation of the geometric series of double-lined chain diagrams depicted in Fig.~\ref{f:squaredcorrelator}, where the lines stand for disorder averaged single propagators
\bea
G_1(r)\equiv \beta\frac{\overline{\langle s(0) s(r)\rangle_c}}{\overline{\Pi}^2}=\int_k \frac{\exp[ikr]}{{\cal J}_k^{-1} +\overline{\Pi}},
\eea
and the local insertions are given by $g \equiv \overline{\Pi^2}-\overline{\Pi}^2$,
\bea
\label{squarecorrelator}
\overline{\langle s(0) s(r)\rangle^2}\sim \int_k \frac{1}{G_2^{-1}(k)-g},
\eea
where $G_2(k)=\int d^3 r \exp[ikr] G_1^2(r)$.
Unlike the sum in Eq.~(\ref{averagepropagator}), this series becomes critical at a finite temperature $T_c$ where
\bea
\label{Tc2}
g\, G_2(k=0)=1.
\eea

We can check that in the large disorder limit this coincides with the critical temperature determined from the instability of the single site model. The variance of the polarizability is dominated by
\bea
g\approx \overline{\Pi^2}\approx \beta^2\int dy \frac{P_{\rm RS}(y)}{\cosh^4(\beta y)}.
\eea
Further, we have
\bea
\label{G2k0}
G_2(k=0)&=&\int d^3r G_1^2(r)=\int_k \frac{{\cal J}^2_k}{(1+{\cal J}_k\overline{\Pi})^2}\nn\\
&=&\frac{2\sqrt{\pi}}{\sqrt{\overline{\Pi}}}\,(1+O(\sqrt{{\overline{\Pi}}})),
\eea
in an expansion for small $\overline{\Pi}$. To this order (which probes the universal properties of the long range Coulomb interactions) this expansion coincides with the expansion for $\psi[(r^{\rm RS}_{\rm sc})^2]\approx \psi[1/\overline{\Pi}]$ (cf., Eq.~\ref{psi2}), while a more careful analysis shows that the two expressions actually agree to higher order than suggested by the leading corrections in (\ref{psi2},\ref{G2k0}). This shows the physical equivalence of the instability of the glass transition (\ref{RSinstab}) and the onset of critical fluctuations in the screening, Eq.~(\ref{Tc2}). As a byproduct, we obtain an interpretation of the function $\psi$ as the long-distance limit of the ladder of doubled propagator lines, which describes the propagation of spin fluctuations.

The criticality of the geometric series for $\overline{\langle s(0) s(r)\rangle^2}-\overline{\langle s(0) s(r)\rangle}^2$ translates into a power law decay of the square of the correlations with distance at the critical point. This critical behavior is the direct analog of the divergence of the non-linear susceptibility in long range spin glasses. In the SK-model with random fields, the dominant part of the non-linear susceptibility is given by the sum $N^{-1}\sum_{ij}\overline{\langle s_i s_j\rangle_c^2}$. It diverges as the Almeida-Thouless instability is approached due to the same kind of criticality as discussed above.

\subsection{Nature of the glass transition}
The Landau expansion for the single site model in terms of fluctuations $\delta \Lambda$ has the same structure as that of the random field SK-model.~\cite{fn3}
For both models the glass transition is continuous in the sense that the effective coupling matrix $\Lambda$, as well as physical properties like the density of states, evolve continuously across the transition. Furthermore, the breaking of ergodicity is weak around $T_c$, the configurational entropy (the logarithm of the number of metastable states) increasing continuously from zero at $T_c$. This is in contrast to structural glasses which usually undergo a discontinuous jamming transition.

\section{The glassy phase}
\label{s:Glassphase}
\subsection{Parisi's Ansatz}
\label{ss:ParisiAnsatz}
Below $T_c$ the replica symmetry of the single site model is broken spontaneously which reflects the occurrence of many metastable states and the associated dynamical slowing down.

In order to describe the low temperature phase, we take Parisi's ultrametric {\em Ansatz} for the pattern of the replica symmetry breaking in the glassy phase~\cite{FisherHertz,MezardParisi87b}. Technically, this amounts to grouping the $n\ra 0$ replicas into sets of $0<x_1<1$, and $x_2/x_1$ ($x_2<x_1$) such sets into a bigger clusters of $x_2$ replicas etc., building an ultrametric tree structure among all replicas. The values assigned to the coupling matrix $\Lambda_{ab}$ are taken to be a decreasing function of the distance on this tree. The coupling matrix may thus be conveniently parametrized as a piecewise constant and increasing function $\Lambda(x)$ with $0<x<1$.
In the present case, the optimal self-consistent solution will turn out to tend to an infinite number of hierarchical subdivisions (continuous replica symmetry breaking) as described by a continuous coupling function $\Lambda(x)$, as in the SK model. Accordingly the overlap between replicas translates into a monotonously increasing function $Q(x)$.

\subsection{Static and dynamic interpretation}
\label{ss:InterpretationofRSB}

Parisi gave an equilibrium interpretation of this Ansatz~\cite{Parisi83}, showing that  $P(q)=dx/dQ|_{Q(x)=q}$ describes the probability of two configurations of the system, sampled with the Boltzmann weight, to have mutual overlap $q$. This interpretation has recently been generalized beyond the mean field case~\cite{FMPP99}.

An alternative interpretation of Parisi's Ansatz which will prove useful for the following, was given by Sompolinsky~\cite{Sompolinsky81} in terms of out-of-equilibrium Langevin dynamics. He supposed the existence a hierarchical set of time scales $t_1\ll t_2\ll...\ll t_k$, all of which are much larger than the microscopic time scale $t_0$. On the level of mean field equations for the dynamics, he {\em assumed} an anomaly in the response at each time scale, modifying the standard response predicted by the fluctuation-response theorem by a multiplicative factor $x(t_i)<1$,
\bea
\label{anomalousresponse}
\d R(t_i)=-\frac{x(t_i)}{T} \d Q(t_i),
\eea
where $R(t)$ is the integrated response function and $Q(t)$ is the local spin-spin correlator averaged over time $t$ and over all sites. The expression $T \d R(t)/\d x (t)$ is sometimes referred to as the response anomaly $\Delta(t)$. Later work~\cite{FranzVirasoro00} has clarified the meaning of this relationship as arising from sampling different metastable states in response to an external field, which is sensitive to the configurational entropy (number of states), rather than the standard entropy, entailing the factor $x$. We will come back to this point in Secs.~\ref{ss:FDT} and \ref{s:Aging} where we will discuss the violations of the fluctuation dissipation theorem and aging.

However, Sompolinsky's theory was criticized because it invoked the divergence of all time scales and their ratios $t_{i+1}/t_i$ with the system size $N$, which is inconsistent~\cite{FisherHertz}. The theory makes sense, however, if we think of the time scales $t_i$ as diverging, e.g., with the waiting time $t_w$ after a sudden quench, rather than with $N$. In the limit where a strong hierarchy of times exists, one can show that the only time dependence of observables is captured by the value of the correlations, or equivalently, by the decreasing function $x(t)$, which thus can serve as a parametrization of time. Remarkably, in the limit of an
infinite number of time scales $k\ra \infty$
the average spin-spin correlation function, $\langle s_i(t)s_i(0)\rangle =Q(x(t))=Q(x)$, of the dynamical formalism turns out to be identical to Parisi's overlap function (cf., Ref.~\onlinecite{CrisantiLeuzzi06} for a recent discussion of this point).

The dynamics described by Sompolinsky's formalism can be understood as a slow relaxation which starts from an out-of-equilibrium configuration which is in general arbitrarily far away from the ground state.~\cite{fn4}
This interpretation is consistent with the picture arising in the mean field aging dynamics as obtained by Cugliandolo and Kurchan.~\cite{CugliandoloKurchan93,CugliandoloKurchan94}

\subsection{Solving the self-consistent equations below $T_c$}
\label{ss:SolutionBelowTc}
In this subsection we show how the self-consistency equations (\ref{SC1}-\ref{SC2}) can be solved under the assumption that the replica matrices $\Lambda_{ab}$, $Q_{ab}$ assume a hierarchical Parisi form.

In order to compute $\langle s_{a}s_{b}\rangle_{H_0}$ in (\ref{SC1}) we follow the scheme introduced by Sommers and Dupont~\cite{SommersDupont84}. We introduce a free energy per replica $\phi(x,y)$ corresponding to the Sompolinsky time scale $x$ (implicitly defined as the time scale over which the system explores metastable states up to an overlap distance $Q(x)$ from the initial state), and in the presence of an external field $y$,
\bea
\label{defofphi}
\exp[x\beta \phi(x,y)]&=& \sum_{s_a=\pm 1 }
\exp\left[ {\cal
H}(x,y,\{s_a\})\right],\\
{\cal H}(x,y,\{s_a\})&=&
\frac{\beta^2}{2}\sum_{a,b=1}^{x} s_a
\Lambda^{(x)}_{ab} s_b +\beta \sum_{a=1}^{x}y s_a.\nn
\eea

Here $\Lambda^{(x)}_{ab}$ denotes the $x \times x$-matrix $\Lambda_{ab}-\Lambda(x)$ with $a,b \in \{1,\dots,x\}$).
The representation (\ref{defofphi}) allows for the derivation of a recursion equation for $\phi(x-dx,y)$
in terms of $\phi(x,y)$ (see, e.g., Ref. \onlinecite{Duplantier81}).
 In the limit of continuous $\Lambda(x)$  and $dx\rightarrow 0$, they reduce to
Parisi's differential equation
\bea &&\frac{\p \phi}{\p x}=-\frac{1}{2}\frac{d \Lambda}{dx} \left[ \frac{\p^2 \phi}{\p y^2} +\beta  x\left(\frac{\p
\phi}{\p y}\right)^2 \right].
\eea
The boundary condition at $x=1$ follows from (\ref{defofphi}) as
\bea
\phi(x=1,y)=\tilde{\Lambda}-\Lambda(1)+\frac{1}{\beta}\log \left[2\cosh(\beta y)\right].
\eea
It is often more convenient to work with the derivative $m(x,y)=\p \phi(x,y)/\p y$,
which satisfies the flow equation
\bea
\label{mdot}
\frac{\p{m}}{\p x}&=&-\frac{1}{2}\frac{d \Lambda}{dx} \left( \frac{\p ^2 m}{\p y^2}+\beta x \frac{\p[m^2]}{\p y}\right),
\end{eqnarray}
and initial conditions on shortest time scales
\bea
\label{bcm}
m(x=1,y)=\tanh(\beta y).
\eea

The above is a priori a formal algebraic trick to compute the free energy density $\phi(0,0)=\lim_{n\rightarrow 0}[\sum_{s_a}\exp[\beta H_0]-1]/n$, from which observables can be obtained by derivatives. However, it can be given a deeper physical meaning in the spirit of Sompolinsky's dynamical interpretation. As shown by Sommers and Dupont~\cite{SommersDupont84}, $m(x,y)$ can be interpreted as the mean occupation (magnetization in spin language) of a site, averaged over the "time scale" $x$ and  in the presence of a local field $y$ which is frozen over that time scale.

\subsection{Distribution of local fields}
\label{ss:Pofy}
We further introduce the distribution of these time-averaged local fields, produced by interactions with other parts of the system and remaining frozen on the "timescale" $x$ due to the glassiness.
This distribution $P(x,y)$ is implicitly defined by the requirement that
\bea
\langle s^{i_1}_{a_1} \dots s^{i_r}_{a_r} \rangle&=&\int dy P(x,y) \langle \left(s^{i_1}_{a_1} \dots
s^{i_r}_{a_r}\right)\rangle _{{\cal H}(x,y)}\nn
\eea
for all
$a_1,\dots,a_r \in \{1,\dots,x\}$.
In the continuous limit, the ensuing recursion relations relating $x$ and $x+dx$ lead to the flow equations
\begin{eqnarray}
\label{Pdot}
\frac{\p P}{\p x}&=&\frac{1}{2}\frac{d \Lambda}{d x} \left(\frac{\p^2 P}{\p y^2}-2\beta x \frac{\p [P\,m]}{\p y}\right).
\eea
The boundary condition for the largest time scales ($x\ra 0$) is
\begin{eqnarray}
\label{bcP}
P(x=0,y)&=&\frac{\exp\{-y^2/2[W^2+\Lambda(0)]\}}{\sqrt{2 \pi [W^2+\Lambda(0)]}},
\eea
which describes the self-consistent Gaussian disorder produced by the bare disorder as well as by Hartree interactions with the frozen part of charges that persists to the longest time scales. As in the replica symmetric case $\Lambda(0)$ does not vanish because of the random field disorder. The same holds true in the random field SK model.

The object of ultimate interest is the distribution of frozen local fields $P(x=1,y)$, the low temperature analog of the distribution (\ref{PRS}). It describes the distribution of thermodynamic fields on shortest time scales (but still $\gg t_0$), over which the system can only explore a single metastable state for lack of time to cross a barrier to another state.

\subsection{Selfconsistency}
\label{ss:selfconsistency}
In order to determine $\Lambda(x)$ self-consistently, we need to calculate the two point correlation function
\begin{eqnarray}
\label{qab}
Q(x)=\langle s_{a}s_{b}\rangle_{H_0,x_{ab}=x} =\int_{-\infty}^{\infty} dy{P}(x,y) m^2(x,y),
\end{eqnarray}
which together with Eq.~(\ref{SC2}) closes the self-consistency loop.

Taking the derivative of Eq.~(\ref{qab}) with respect to $x$ and using  the flow equations (\ref{mdot},\ref{Pdot}), we obtain the more convenient form
\begin{eqnarray}
\label{qdot}
\frac{d Q(x)}{d x}=\frac{d {\Lambda(x)}}{d x}\int_{-\infty}^{\infty} dy {P}(x,y) \left(\frac{\p m(x,y)}{\p y}\right)^2.
\end{eqnarray}

The self-consistency equations can be solved iteratively.
At the $n^{\rm th}$ stage, the iteration loop consists in computing
\be
{\bf Q}_{n}={\bf Q}_{n}(\{{\bf \Lambda}_n\})
\label{iter1}
\ee
from (\ref{qab}), and a new coupling matrix via
\begin{equation}
\beta{\bf \Lambda}_{n+1}=\chi_1^{-1}(\beta {\bf Q}_n)-1/\beta{\bf Q}_n.
\label{iter2}
\end{equation}
The latter can be done conveniently in replica Fourier space, see App.~\ref{app:RFT}. For some details of the numerical implementation we refer the reader to App.~\ref{app:numerics}.

\subsection{Marginal stability of the glass phase}
\label{ss:marginalstability}
In this section, we show that throughout the low temperature phase the system is in a marginal state, provided that the replica symmetry is continuously broken, or more precisely, provided that $d{\Lambda}/dx$ is continuous and bounded as $x\ra 1$. In other words, permanent marginal stability is guaranteed by the absence of step-like discontinuities in $\Lambda(x)$ on the shortest time scales. This assumption is confirmed by the full numerical solution discussed in the following section.~\cite{fn5}

Let us rewrite the self-consistency equation (\ref{SC2}) using the Fourier transform identity (\ref{RFT2}) as
\begin{eqnarray}
\label{SCrew}
\beta (Q_c-[Q](x))&=&\chi_1\left[ r_{\rm sc}^2(x)\right],
\end{eqnarray}
where
\begin{eqnarray}
\left[Q\right](x)&\equiv &x Q(x)-\int_{0}^x Q(y) dy,\\
Q_c&\equiv & \tilde{Q}-\int_0^1 Q(x) dx.
\end{eqnarray}
We have also used the shorthand notation
\begin{eqnarray}
\label{rsc}
r_{\rm sc}^2(x)=\beta^2(\Lambda_c-[\Lambda](x))+1/(Q_c-[Q](x)),
\end{eqnarray}
recalling that $r_{\rm sc}(x)$ takes the place of an (average) Thomas-Fermi screening length on the time scale $x$. (In technical terms, $r_{\rm sc}^2$ is the replica Fourier transform of the inverse polarizability operator $-\Pi^{-1}$, cf.~(\ref{SP}).) However, the term "screening length" should not be taken too literally, since we have seen that homogeneous screening breaks down below $T_c$. As we will see below $r_{\rm sc}$ decreases with increasing time (decreasing $x$), reflecting that on long time scales the electron glass screens better on average.

Taking the derivative of Eq.~(\ref{SCrew}) with respect to $x$, we find after a little algebra
\begin{eqnarray}
\label{dotratio}
\frac{\d Q(x)/\d x}{\d \Lambda/ \d x}
= \frac{1}{\psi[r_{\rm sc}^2(x)]}.
\end{eqnarray}
Combining with Eq.~(\ref{qdot}) we find the relation
\begin{eqnarray}
\label{SCderived}
\frac{1}{\psi[r_{\rm sc}^2(x)]}=\int_{-\infty}^{\infty} dy {P}(x,y) \left[\frac{\p m(x,y)}{\p y}\right]^2,
\end{eqnarray}
which expresses the marginality of the full RSB solution at all scales $x$ where $\Lambda(x)$ is continuous.

Evaluating it for $x=1$, we find in particular
 \begin{eqnarray}
\label{marginalstability}
\frac{1}{\psi[r_{\rm sc}^2(1)]}=\int_{-\infty}^{\infty} dy \frac{\beta^2 P(1,y)}{\cosh^4(\beta y)},
\end{eqnarray}
extending the relation (\ref{RSinstab}) from $T_c$ to the low temperature regime.

As mentioned above, Eq.~(\ref{marginalstability}) is equivalent to the condition for marginal stability on shortest time scales, or in technical terms, to the marginal stability of the free energy  (\ref{freeenergy}) with respect to replicon fluctuations~\cite{MuellerIoffe04}. Note that the above reasoning shows that any model with continuous replica symmetry breaking is marginally stable throughout the low temperature phase, as was derived in a different way in Ref.~\onlinecite{CLPR03}, too.

Besides Eq.~(\ref{marginalstability}), an additional relation between the screening radius  $r_{\rm sc}(1)$ and the field distribution $P(1,y)$ follows from the self-consistency of the spin-spin correlator, i.e., the compressibility Eq.~(\ref{SCrew}),
\begin{eqnarray}
\label{susclowT}
\kappa_C =\frac{d n}{d\mu}&\equiv &\chi_1[r_{\rm sc}^2(1)]=\beta(Q_c-[Q](1))\nn\\
&=&\beta(1-Q(1))=\int_{-\infty}^{\infty} dy \frac{\beta P(1,y)}{\cosh^2(\beta y)}.
\end{eqnarray}

Anticipating a large screening radius $r_{\rm sc}(1)$ at low temperature, we can use the asymptotic expansion of $\chi_1$, Eq.~(\ref{fasym}), and $\psi$, Eq.~(\ref{psi}), to conclude that
\begin{eqnarray}
\int_{-\infty}^{\infty} dy \frac{\beta P(1,y)}{\cosh^2(\beta y)}&\approx& \frac{1}{r^2_{\rm sc}(1)},\\
\int_{-\infty}^{\infty} dy \frac{\beta^2P(1,y)}{\cosh^4(\beta y)}&\approx&\frac{1}{\sqrt{\pi}r_{\rm sc}(1)}.
\end{eqnarray}

A consistent and regular low temperature solution of both equations requires a screening radius $r_{\rm sc}(1)\sim 1/T$ and a scaling form of the frozen field distribution in metastable states
\bea
\label{Pscaling}
P(1,y)=T^2\Psi_{\rm 3D}(y/T),
\eea
with $\Psi_{\rm 3D}(\epsilon)\sim \epsilon^2$ for $\epsilon \gg 1$.
This will indeed be confirmed by the full solution in the next section.

Note that this scaling implies $P(y)\sim y^2$ for $y\gg T$, which is exactly Efros and Shklovskii's prediction for the Coulomb gap in 3D. Simple arguments for Thomas Fermi screening in presence of this pseudogap also suggest an average screening on the scale $r_{\rm sc}\sim 1/T$ where thermal fluctuations and Coulomb interactions compete.

Interestingly, the saturation of the pseudogap exponent $2=D-1$ comes out as a necessary consequence of the marginal stability condition, without requiring any further assumption. We may interpret this result as follows. The fact that the dominant metastable states of the Coulomb glass are just marginally stable means that the system assumes the highest local density of states which is still just compatible with stability requirements. The marginal states exhaust all these degrees of freedom, which leads to the saturation of the Efros-Shklovskii exponent.

This result is interesting not only because it gives a justification of the saturation, but also because the present theory takes into account the collective response of many neighbors to the presence of a given electron, and thus goes beyond a stability analysis with respect to single particle hops. Our result shows that such many particle effects do in fact not alter the gap exponent. However, we will see that it suppresses the coefficient of the power law describing the pseudogap in the density of states.

Note that the exponent being $2$ is not merely a coincidence. In fact, in arbitrary dimensions $D>2$ with $J(r)\sim 1/r^\nu$ and $J(k)\sim 1/k^{D-\nu}$, an analysis along the same lines leads to $P(1,y)=T^{D/\nu-1}\Psi_{D,\nu}(y/T)$, with $\Psi_{D,\nu}(\epsilon)\sim \epsilon^{D/\nu-1}$ for $\epsilon \gg 1$, again in agreement with the stability arguments by Efros and Shklovskii. In the two-dimensional case with $1/r$ interactions, there are logarithmic corrections (see the remark after Eq.~(\ref{psi})).

The physical implications of the marginality of the glass will be discussed in Sec.~\ref{s:Discussion}

\section{Low temperature analysis}
\label{s:lowTanalysis}
Let us now analyze the low temperature behavior of the mean field equations in more detail.

In order to solve the Parisi equations in the low temperature limit we introduce the rescaled variable $b\equiv x/T$ which appears in the anomalous response (\ref{anomalousresponse})
as the inverse of an "effective temperature" $T/x$, replacing the real temperature for equilibrium response. Indeed, in Sec.~\ref{ss:FDT} we will review that the relation~(\ref{anomalousresponse}) is actually just the generalized fluctuation-dissipation relation in disguise which is usually taken as a definition of "effective temperature".

We anticipate the most important variations of $Q(x)$ and $\Lambda(x)$ on the scale of $x\sim T$, as is usually the case in mean field glasses~\cite{VannimenusToulouse81}.
Such a scaling has been observed in the recent numerical solution of the SK model at low temperature~\cite{CrisantiRizzo02}, and has been exploited later to resolve the asymptotic $T\rightarrow 0$ limit~\cite{OppermannSherrington05,Pankov06}.

The analysis in the previous section suggests that at low $T$ an Efros-Shklovskii Coulomb gap develops, $P(x=1,y)\sim y^2$, and accordingly we expect typical overlaps $1-Q\sim T^3$ and screening radii $r_{\rm sc}\sim 1/T$. We will see below that this scaling not only holds on shortest time scales $x=1$, but extends to larger time scales, too, provided the temperature is replaced by the "effective temperature" $T/x=1/b$. In particular, we will find the overlap $1-Q(x)\sim 1/b^3$ and the average screening radius $r_{\rm sc}(x;T)\sim b$ to decrease with increasing effective temperature, or equivalently, with increasing time.

\subsection{Regular flow equations}
\label{ss:lowT}
The effective temperature sets a characteristic energy scale at time scale $x$. This suggests to rescale the local fields $y$ by $T/x$, as was done in Ref.~\onlinecite{Pankov06} for the SK model. We thus change variables to $z\equiv f(b)y\equiv (c+b) y$, where an arbitrary constant $c=O(1)$ is included to make the variable change regular for $b=\beta x\rightarrow 0$. (When plotting results for $p(z,b=\beta)$ later, we will always revert to $c=0$, however).
 Anticipating an Efros-Shklovskii-pseudogap $P(y)\sim y^2$, we further introduce the rescaled field distribution
\bea
\label{p_def}
p(b,z,T)&=&b^2 P(x=bT,y=z/f(b)),
\end{eqnarray}
for which we expect $p(z)\sim z^2$ at large $b=\beta x$ and $z\gg 1$.

In terms of the variables $b$ and $z$, the magnetization and the rescaled field distribution obey the flow equations
\begin{eqnarray}
\label{smeq}
 f(b)\frac{\p m}{\p b}&=&-z f'(b)\frac{\p m}{\p z}\\
 &&\quad -\frac{b f^2(b)}{2}\frac{d{ \Lambda}}{d b}\left(\frac{\p^2 m}{\p z^2}+\frac{b}{f(b)}\frac{\p\left[m^2\right]}{\p z}\right),\nn
\\
\label{speq}
 f(b)\frac{\p p}{\p b}&=&-z f'(b)\frac{\p p}{\p z}+2 p\\
 &&\quad +\frac{b f^2(b)}{2}\frac{d{ \Lambda}}{d b}\left(\frac{\p^2 p}{\p z^2}-2\frac{b}{f(b)}\frac{\p [p m]}{\p z}\right),\nn
\end{eqnarray}
with  boundary conditions that remain regular in the limit $T\ra 0$,
\begin{eqnarray}
\label{bclowT}
m(b=\beta,z)&=&\tanh[z f(\beta)/\beta],\\
p(b=0,z)&=&\frac{\exp\{-[z/f(0)]^2/2[W^2+\Lambda(0)]\}}{f^2(0)\sqrt{2 \pi [W^2+\Lambda(0)]}}.
\end{eqnarray}
The virtue of the rescalings is that $\beta$ has disappeared from the flow equations, and only remains present in the boundary conditions for $m$. In the new variables the
low temperature limit is completely regular, and focuses on the features of the physically active degrees of freedom described by $z=O(1)$. This will be exploited in the next section.

The local spin-spin correlator (overlap) is obtained from the relations
\begin{eqnarray}
\label{lowTQ}
&&1-Q(b=\beta)=\left.\frac{1}{b^2 f(b)}\int_{-\infty}^{\infty} dz\, p(b,z) [1-m(b,z)^2]\right|_{b=\beta}\nn\\
&&\hspace{2cm}=\frac{1}{\beta^2 f(\beta)}\int_{-\infty}^{\infty} dz\, \frac{p(b=\beta,z)}{\cosh^2(z)},\\
\label{QdotlowT}
 &&\frac{d Q(b)}{d b}=\frac{1}{f(b)}\frac{d \Lambda(b)}{d b} \int_{-\infty}^{\infty} dz\, p(b,z) \left[\frac{\p m(b,z)}{\p z}\right]^2.
\end{eqnarray}
From the first equation we expect $1-Q(b)\sim b^{-3}$. If we further assume little dependence on $\beta$ at a fixed $b$ we expect $d Q/d b\sim b^{-4}$, which together with the second relation suggests $d \Lambda/d b \sim b^{-3}$. This scaling will indeed be confirmed below at low but finite temperature, and will be used for the strict $T=0$ limit in Sec.~\ref{s:fixedpoint}.

The self-consistency condition Eq.~(\ref{SCrew}), and its first derivative with respect to $b=x/T$, read
\begin{eqnarray}
\label{SClowT}
\chi_1[r_{\rm sc}^2(b=\beta)]&=&\beta[1-Q(b=\beta)],\\
\frac{d [r_{\rm sc}^2(b)]}{d b} \chi_1^\prime[r_{\rm sc}^2(b)]&=& -b\frac{d Q(b)}{d b},\label{SClowT2}
\end{eqnarray}
which allows us to obtain the average screening radius $r_{\rm sc}(b)$ from the  overlap function $Q(b)$.
We will show in Sec.~\ref{ss:FDT}, that Eq.~(\ref{SClowT2}) simply expresses Sompolinsky's anomalous response relation (\ref{anomalousresponse}) at scale $x=bT$.

Finally, we re-compute the coupling matrix $\Lambda(b)$ by integrating the equations
\begin{eqnarray}
\label{LambdaT1}
\Lambda(b=0)&=&Q(b=0)\psi[r_{\rm sc}^2(0)],\\
\label{LambdaT2}
\frac{d \Lambda(b)}{d b}&=&\frac{d Q(b)}{d b}\psi[r_{\rm sc}^2(b)],
\end{eqnarray}
which closes the self-consistency loop $\Lambda(b)\rightarrow Q(b) \rightarrow r^2_{\rm sc}(b)\rightarrow \Lambda(b)$.

\subsection{Results for the glass phase}
\label{ss:resultsatlowT}
We have solved the self-consistency loop (\ref{smeq}-\ref{LambdaT2}) numerically for temperatures down to $T=0.01$, both for moderate and relatively strong disorder, $W=2,10$. Details of the implementation are given in App.~\ref{app:numerics}.

Our results confirm that the glass transition occurs in a continuous fashion at $T_c$, excluding a discontinuous breaking of replica symmetry, as is consistent with a stability analysis in a Landau expansion around $T_c$. The coupling function $\Lambda(x)$ was found to remain regular everywhere, ensuring marginality at all scales.

\subsubsection{Opening of the pseudogap}

In Figs.~\ref{f:PhPyw=2}, \ref{f:PhPyw=10} the temperature evolution of the pseudogap below the glass transition is shown. Apart from the overall width of the field distributions, moderate and strong disorder ($W=2,10$) do not differ greatly.
Notice the fast opening of the pseudogap in the distribution of the thermodynamic fields. This translates into a fast drop of the compressibility below $T_c$. It may also affect transport properties to the extent that $P(y)$ contains information about the multiparticle excitations (electronic polarons) which are believed to carry hopping current~\cite{ESbook}.

On the other hand, the distribution of instantaneous fields displays a prominent plasma dip already at temperatures substantially above $T_c$, as shown in Fig.~\ref{f:PhhighT}. The further suppression of this dip affects mostly very low energies.

Both for $P(h)$ and $P(y)$ only that low energy regime is expected to exhibit a universal ES pseudogap. For better comparison of the temperature evolution of the two distributions, we display them together in Fig.~\ref{f:Pyandh_w=2} for $T\gg T_c$, $T\approx T_c$ and $T\ll T_c$.

\begin{figure}
\includegraphics[width=3.0in]{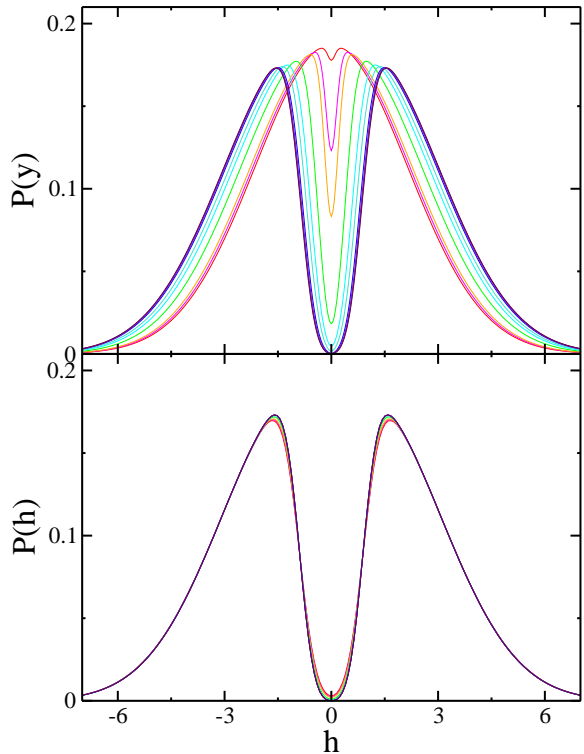}
\vspace{.15cm}
\caption{(Color online) Evolution of the pseudogap in the distribution of the thermodynamic fields $P(y)$ (top panel) and the instantaneous fields $P(h)$ (bottom panel) for moderate disorder $W=2$. The curves are plotted for the
inverse temperatures $\beta=7.2, 7.6, 8, 10, 14, 20, 40, 60, 80, 100$ (from top to bottom) below the glass transition ($\beta_c = 7.149$). Note the rapid opening of the pseudogap in $P(y)$ just below the transition. Most of the suppression of the instantaneous density of states is already formed as a plasma dip at the transition.}
\label{f:PhPyw=2}
\end{figure}

\begin{figure}
\includegraphics[width=3.0in]{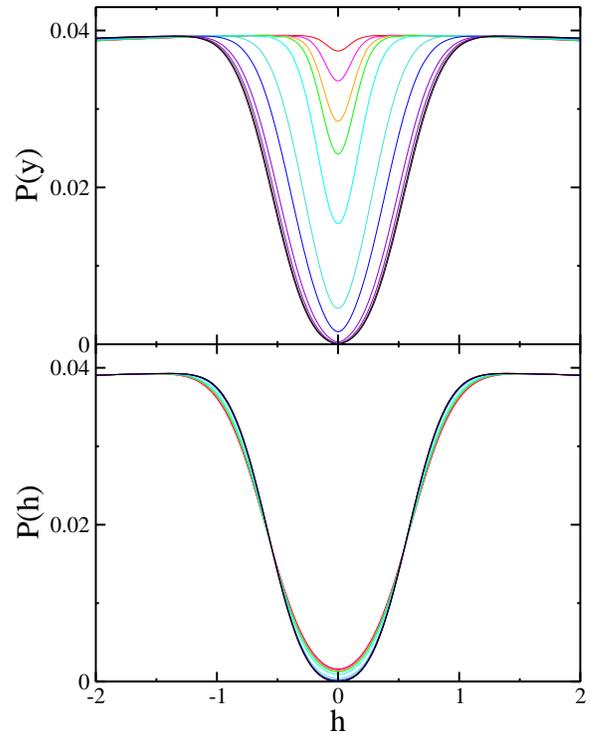}
\vspace{.1cm}
\caption{(Color online) Evolution of the gap in the distribution of the thermodynamic fields $P(y)$ (top panel) and the instantaneous fields $P(h)$ (bottom panel) in strong disorder $W=10$. The curves correspond to
inverse temperatures $\beta = 8.2, 8.4, 8.7, 9, 10, 14, 20, 40, 60, 80, 100$ below the glass transition ($\beta_c\approx 8.134$). The salient features are similar as for moderate disorder, cf., Fig.~\ref{f:PhPyw=2}.}
\label{f:PhPyw=10}
\end{figure}

\begin{figure}
\vspace{.2cm}
\includegraphics[width=3.0in]{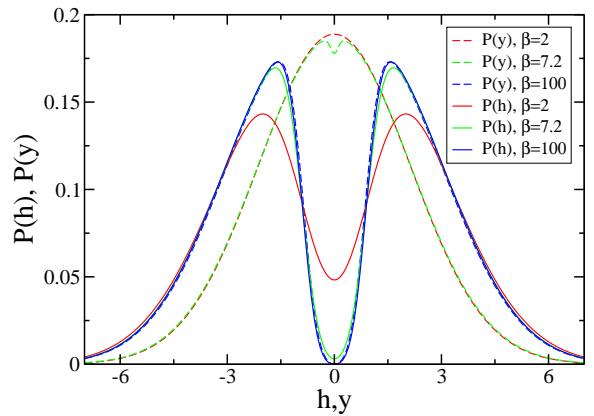}
\caption{(Color online) The distribution of instantaneous fields $h$ (solid lines) and thermodynamic fields $y$ (dashed lines) for temperatures well above ($\beta=2$, red), close to ($\beta=7.2$, green) and well below ($\beta=100$, blue) the glass transition at moderate disorder $W=2$. Notice the presence of a pronounced plasma dip in $P(h)$ even at high temperature.}
\label{f:Pyandh_w=2}
\end{figure}

\subsubsection{Compressibility}
In the main panel of Fig.~\ref{f:Itunn_w=2_compare} we plot the compressibility (\ref{susclowT}) as function of $T$. The higher order non-analyticity at the transition temperature $T_c$ is hard to discern. It would in principle show up more markedly in a plot of the non-linear capacitance, which is the analog of the non-linear susceptibility of spin glasses.

Similarly as in spin glasses, the low temperature, zero-field cooled compressibility  is strongly suppressed as
\bea
\label{ZFC}
\kappa_C\equiv \kappa_C^{\rm ZFC}=\beta[1-Q(1)] \sim T^2,
\eea
reflecting the lack of short-time ergodicity of the system. On the other hand, the equilibrium, or field cooled, prediction for the compressibility is
\bea
\kappa_C^{\rm FC}=\beta\int_0^1 dx\, [1-Q(x)]\stackrel{T\ra 0}{\ra} {\rm const}.
\eea
This observable does not feel the Coulomb gap and is thus much less suppressed. The field-cooled compressibility is only accessible by raising the temperature back above $T_c$ and quenching the system in the new field, or alternatively, in the ultra-long time limit at a constant low temperature.
The unstable, replica-symmetric high temperature solution similarly predicts a constant susceptibility as $T\ra 0$.

\subsubsection{Tunneling}
Let us briefly discuss tunneling from a broad junction into a strongly insulating classical electron glasses, as studied experimentally in Ref.~\onlinecite{massey95}. The broad junction assures self-averaging of the tunneling current. In the insulating limit, and for voltages well below the temperature, $eV < k_B T$, the tunneling current is dominated by the hopping of electrons from the probe to empty sites in the electron glass. Such hops are usually not accompanied by any further rearrangements of other electrons, since those are exponentially suppressed in terms of the action.

In contrast to the compressibility, tunneling is thus controlled by the distribution of instantaneous fields, $P(h)$, (\ref{PofhfromPofy}). In the linear response regime the differential tunneling conductance is proportional to
\be
\label{Gtunn}
G(T)\equiv \frac{d I}{d V}(V\ra 0)\propto \int dh \frac{P(h;T)}{\cosh^2(\beta h)},
\ee
which is plotted in the inset of Figs.~\ref{f:Itunn_w=2_compare}.

\begin{figure}
\vspace{.2cm}
\includegraphics[width=3.0in]{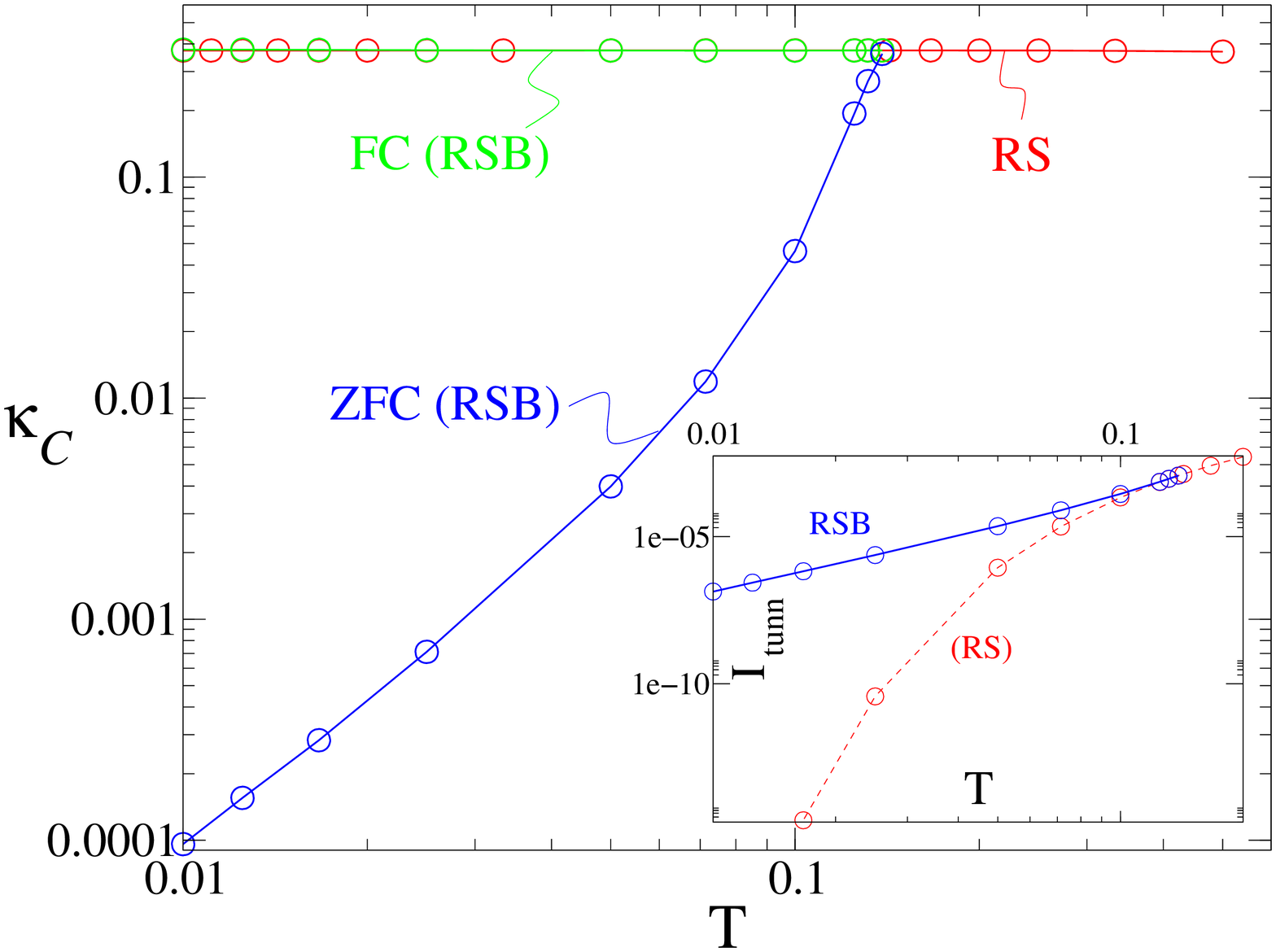}
\caption{(Color online) Main panel: Zero-field cooled compressibility $\kappa_C$ (ZFC, blue line), equilibrium field cooled compressibility (FC, green line) and the replica symmetric result (RS, red line) for $W=2$. The extrapolation of the RS prediction below $T_c$ is unphysical, but closely follows $\kappa^{\rm FC}_C$, as one may expect. Inset: Tunneling conductance $G$ as predicted by Eq.~(\ref{Gtunn}) in the glass phase (RSB). The unphysical replica symmetric theory (RS) would predict a strong suppression of $G$ due to a constant Onsager term and an ensuing hard gap in $P_{\rm RS}(h)$.}
\label{f:Itunn_w=2_compare}
\end{figure}

\subsection{Scaling and universality}
\label{ss:scaling}
The low temperature solution confirms the anticipated scalings $d Q/d b\sim b^{-4}$, $d \Lambda/d b \sim b^{-3}$ and $r_{\rm sc}\sim b$ for large $b$, $1\ll b \lesssim \beta$. In this regime these functions, at fixed $b$, are nearly independent of temperature and disorder, as illustrated in
Figs.~\ref{f:bindependence} and \ref{f:universality}. As a result, physical quantities such as the compressibility, or the density of states of thermally active degrees of freedom turn out to be remarkably universal, that is, independent of disorder and obeying simple scaling laws as a function of temperature.

\begin{figure}
\vspace{.3cm}
\includegraphics[width=3.0in]{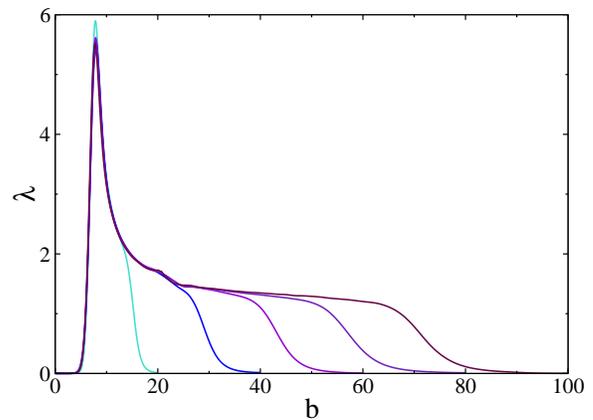}
\caption{(Color online) The coupling function $\lambda(b)$ for $\beta=20,40,60,80,100$, and disorder $W=2$. The result is nearly independent of $\beta$ as long as $\beta_c\ll b\lesssim 0.7 \beta$.}
\label{f:bindependence}
\end{figure}

\begin{figure}
\includegraphics[width=3.0in]{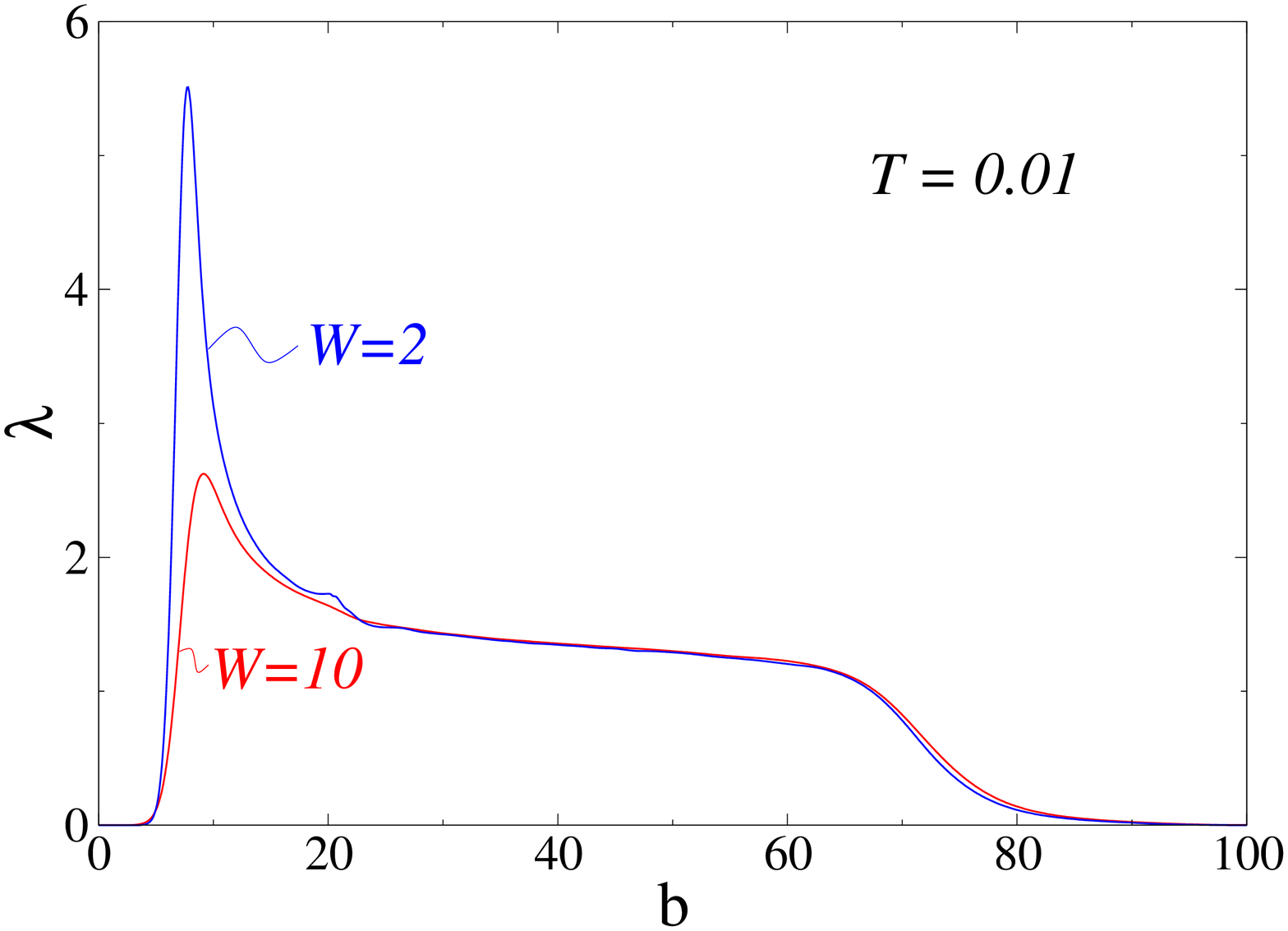}
\caption{(Color online) The coupling function $\lambda(b)$ at low temperature $T=0.01$, for different disorders, $W=2,10$. In the regime $b\gg \beta_c$, the coupling function is universal, i.e., essentially independent of the bare disorder.}
\label{f:universality}
\end{figure}

Furthermore, we find the scaled field distribution function $p(b=\beta,z)$ to approach an essentially temperature independent form at low $T$ (see the top panel of Fig.~\ref{f:Pyscaling}).
This confirms our prediction (\ref{Pscaling}) for the scaling form of $P(x=1,y)$. 

The reason for the asymptotic universality at low $T$ lies in the appearance of an attractive fixed point of the flow equations, as discussed in the next section.

\subsection{Break point and plateau}
On short timescales ($x\ra 1$, $b\rightarrow \beta$), the sole dependence on $x/T$ is violated. The derivative of the coupling function, $b^3d \Lambda/d b$, drops to $0$ rather sharply around $x_P(T)\equiv b_P(T)/\beta <1$, implying
that both $\Lambda(x)$ and $Q(x)$ are constant beyond $x_P(T)$. This plateau in the coupling and overlap function appears below $T_c$, with $x_P(T)$ decreasing to a saturation value $x_P(T\ra 0)\approx 0.78$ at sufficiently low temperature. The low temperature limit of this feature and its physical implications will be analyzed in the following two sections.

\section{Low temperature fixed point}
\label{s:fixedpoint}
\subsection{Flow equations for $T=0$}
Let us now analyze the flow equations in the regime of very low effective temperatures, $T/x=1/b \ll T_c$. In particular we will be interested in the zero temperature limit at finite $x$ where $x\gg T/T_c$ is satisfied.
For this analysis we can use the asymptotic low energy scalings of Eqs.~(\ref{fasym},\ref{psi2})
\begin{equation}
\chi_1(r_{\rm sc}^2)\approx 1/r_{\rm sc}^2,\quad \psi(r_{\rm sc}^2)\approx A[r_{\rm sc}(x)]^{2-2/{\alpha}},
\end{equation}
where we allowed for a more general asymptotics for $\psi$ so as to cover arbitrary space dimensions $D$, with power law repulsions $1/r^\nu$. Note that the parameters $A$ and $\alpha$  derive from the low energy asymptotics of the interactions $J_{ij}$, and are thus independent of lattice details. For Coulomb glasses ($\nu=1$), $\alpha$ is the Efros-Shklovskii exponent $\alpha=D-1$, and in particular $A_{3D}=\sqrt{\pi}$, cf., Eq.~(\ref{psi2}).

The above asymptotic forms simplify the self-consistency equations (\ref{SClowT}-\ref{LambdaT2}) to
\begin{eqnarray}
&&\dot\Lambda(x)=A[r_{\rm sc}(x)]^{2-2/{\alpha}}\dot Q(x),
\label{sceq1}\\
&&\frac{\dot{[r^2_{\rm sc}]}(x)}{r^4_{\rm sc}(x)}= \beta x\dot Q(x),
\label{sceq2}
\end{eqnarray}
where dots denote derivatives with respect to $x$.

The low temperature analysis of the SK model~\cite{Pankov06} leads to completely analogous equations, the only trace of its specific interactions being the values $A=1$, $\alpha=1$ (since $\psi_{\rm SK}\equiv 1$).
The low energy asymptotics of the SK model is thus very similar to a 2D electron glass, apart from logarithmic corrections in the latter which lead to $\psi_{\rm 2D}[r_{\rm sc}^2]\approx 2\pi\ln{[r_{\rm sc}(x)^2]}$.

In order to obtain a regular low temperature limit of the flow equations, we rescale the physical
quantities similarly as in the previous section (without changing the variables $x\ra b$, however),
\bea
z&\equiv& \beta x y,\\
m(x,y) &\to& m(x,z),\\
P(x,y)&\to&(\beta x)^{-\eta} p(x,z),\\
\dot \Lambda(x)&\to&\delta\beta (\beta x)^{-\delta-1} \lambda(x),\\
\dot Q(x)&\to&\zeta\beta (\beta x)^{-\zeta-1} q(x), \label{qvsQdot}\\
1/r_{\rm sc}^2(x)&\to& (\beta x)^{-\rho} \xi(x).
\eea
The exponents $\eta, \delta, \zeta, \rho$ are chosen in
such a way that any explicit singular dependence on $\beta$ is eliminated from the equations.
The form of the flow equations for $m$ and $P$ 
and Eqs.~(\ref{sceq1},\ref{sceq2})
together require 
\bea
\delta &=&2,\\
\eta=\rho=\zeta-1=\alpha&=&\left\{ \begin{array}{cc} D-1 & {\rm Electron\, glass,}\\ 1 & {\rm SK model.}\end{array}\right.
\label{exponents}
\eea
With these notations the flow equations become:
\begin{eqnarray}
&& x\dot{m}=
-\lambda(x)\left(m''+2 m m'\right)-z m',
\label{smeq1}\\
&& x\dot{p}=
\lambda(x)\left(p''-2(p m)'\right)-z p'+\alpha p,
\label{speq1}
\end{eqnarray}
where primes denote partial derivatives $\p/\p z$.
The self-consistency equations~(\ref{sceq1}-\ref{sceq2}) take the form
\begin{eqnarray}
\lambda(x)&=&\frac{A}{2}(1+\alpha)q(x)\xi^{(1/\alpha-1)}(x),
\label{ssceq1}\\
x\dot\xi(x)&=&\alpha\xi(x)-(1+\alpha)q(x),
\label{ssceq2}
\end{eqnarray}
while Eqs.~(\ref{lowTQ},\ref{QdotlowT}) translate into
\bea
\zeta q(x)&=&\delta \lambda(x) \int dz \,p m^{\prime2},\\
\label{qpm}
\xi(x=1)&=&\int dz\, \frac{p(1,z)}{\cosh^2(z)}.
\label{qpm1}
\eea
Notice that for the SK model ($\alpha=1$) Eq.~(\ref{ssceq1}) expresses simply the identity $\lambda(x)=q(x)$, while Eq.~(\ref{ssceq2}) is not needed to obtain a  self-consistent solution.

\subsection{Fixed point}
The system of Eqs.~(\ref{smeq1},\ref{speq1}) looks like a set of functional renormalization
group (FRG) equations as they appear in studies of vortex lattices, magnetic interfaces and wetting problems in the presence of randomness~\cite{Fisher8586,LeDoussal}. In the present case, the "time variable" $x$ takes the role of the renormalization scale, $m$ and $p$ being analogous to functional running coupling constants. However, a formal difference with standard FRG equations consists in the explicit $x$-dependence of the "beta-function" through $\lambda(x)$ which is itself determined by the flow. If $\lambda(x)$ were constant, the flow equations would constitute an autonomous system of partial differential equations (i.e., with a right hand side independent of $x$). Such systems commonly possess fixed points, such as many standard renormalization group flows.

It turns out that despite the remaining $x$-dependence the flow is attracted to an asymptotic fixed point in the intermediate regime $1\gg x\gg T/T_c$, which controls universal low energy physics in a similar manner as in standard renormalization group flows. The flow of the magnetization and field distribution functions to respective fixed points is illustrated in Figs.~\ref{f:mscaling} and ~\ref{f:pscaling}.

The attractiveness of the fixed point wipes out the trace of the bare disorder $W$ entering the flow equations via the boundary conditions at $x=0$, that is, at high effective temperature and energy scales. The flow to lower effective temperatures renormalizes the field distribution $p(z)$ to a universal form, at least for fields describing active degrees of freedom, $y\sim T_{\rm eff}$ or $z=O(1)$). On the other hand, the high energy tails in the unscaled distribution $P(y)$ basically retain the structure set by the bare disorder.

Despite the above suggestive similarities with genuine functional renormalization group flows, we do not have - at present - a thorough understanding of this analogy, even though we believe that there is a deeper connection with FRG approaches.

\begin{figure}
\includegraphics[width=3.0in]{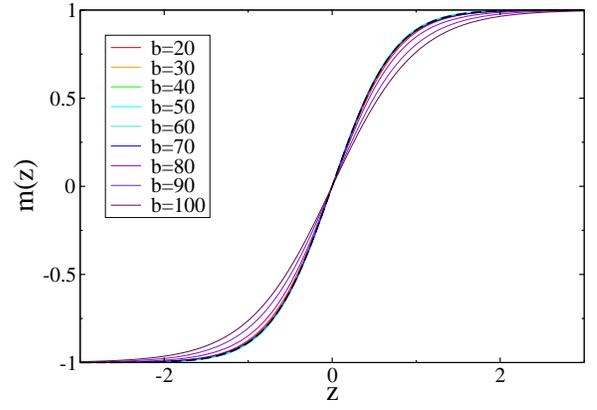}
\caption{(Color online) Illustration of the temporal flow of $m(x,z)$ for increasing Sompolinsky times $x=b/\beta=1,0.9,0.8,\dots 0.2$ for $\beta=100$ and $W=2$. An intermediate fixed point is clearly approached near $x=0.5$. The asymptotic prediction for the fixed point is plotted as a dashed line (black), for which we used an artificial power law fitting $\psi_{\rm 3D}$ beyond its universal low energy regime.}
\label{f:mscaling}
\end{figure}

\begin{figure}
\vspace{.6cm}
\includegraphics[width=3.0in]{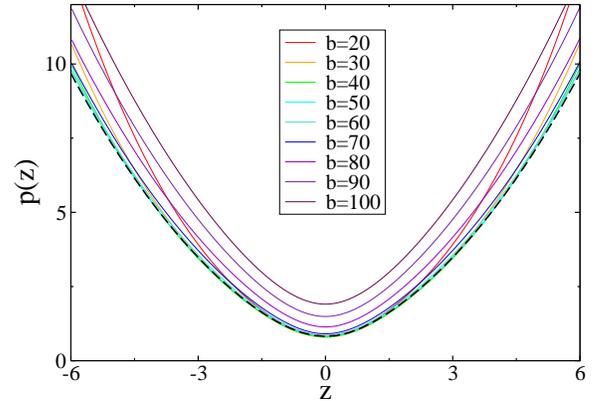}
\caption{(Color online) Illustration of the temporal flow of $p(x,z)$ for increasing Sompolinsky times $x=b/\beta=1.0, 0.9,\dots 0.2$ for $\beta=100$ and $W=2$. The apparent fixed point occurring around $x=0.5$ comes close to the fixed point predicted with an artificial power law which fits $\psi_{\rm 3D}$ beyond its universal low energy regime  (dashed black line). The remaining small discrepancy is due to the approximate nature of the power law fit.}
\label{f:pscaling}
\end{figure}

On the putative fixed point, the derivatives $x\frac{d}{dx}$ vanish. The fixed point functions $m^\star, p^\star$ depend only on the variable $z$, satisfying the equations
\begin{eqnarray}
&&\lambda^\star\left({m^\star}''+2 m^\star {m^\star}'\right)
+z {m^\star}'=0,
\label{fpmeq}\\
&& \lambda^\star\left({p^\star}''-2(p^\star m^\star)'\right)
-z {p^\star}'+\alpha p^\star=0,
\label{fppeq}\\
&&\zeta q^\star= \delta \lambda^\star \int dz\, p^\star (m^{\star\prime})^2,\\
&&\alpha \xi^\star= (1+\alpha) q^\star,
\end{eqnarray}
with the self-consistent value for $\lambda^\star$,
\bea
\label{SC_FP}
\lambda^\star&=&\frac{A(1+\alpha)}{2}q^\star \xi^{\star (1/\alpha-1)}\\
&=&\frac{A\alpha}{2}\left(\frac{1+\alpha}{\alpha}q^\star\right)^{1/\alpha}.\nn
\eea

The fixed point functions are subject to the boundary conditions
\begin{eqnarray}
m^\star(0)={p^\star}'(0)&=& 0,\\
m^\star(z\to\infty)&\to&{\rm sign}(z),\label{icm}\\
p^\star(z\to\infty) &\propto & H_{\alpha}
\left(\sqrt{2\lambda^\star}+\frac{|z|}{\sqrt{2\lambda^\star}}\right),
\label{icp}
\end{eqnarray}
where $H_{\alpha}$ is the Hermite polynomial of order $\alpha$. The last condition
follows from the solution
of the linear Eq.~(\ref{fppeq}) at large $z$ where $m^\star={\rm sign}(z)$. This leaves open a constant of proportionality which has to be fixed eventually by the self-consistency constraint (\ref{SC_FP}).

Eqs.~(\ref{fpmeq},\ref{fppeq}) constitute a self-consistency problem for $\lambda^\star$. They have a unique solution which can easily be found numerically.
For the 3D Coulomb glass we find the fixed point values $\lambda^\star=1.1230287...$, $q^\star=0.26763357...$, $\xi^\star=0.40145035...$. For comparison, in the SK model one finds $\lambda^\star=q^\star=.4108021...$ and $\xi^\star=0.8216042...$.

\subsection{Stability}
The analysis and even the meaning of the attractiveness of the fixed point is not as straightforward as in standard renormalization flows, because of the non-local coupling due to the dependence of $\lambda(x)$ on the flow itself.
However, we can analyze deviations from the fixed point of the form
\bea
m(x,z)&=&m^{\star}(z)+m_1(z)x^{\nu},\\
p(x,z)&=&p^{\star}(z)+p_1(z)x^{\nu},\\
\lambda(x)&=&\lambda^{\star}+\lambda_1 x^{\nu},
\eea
such that the consistency equations (\ref{ssceq1}-\ref{qpm}) are still obeyed.
The functions $m_1$, $p_1$ and $\lambda_1$ will in general depend on the exponent $\nu$.
We found that both for the 3D Coulomb glass and the SK model such linearized deviations require a positive $\nu\geq \nu_{\rm min}>0$. This suggests that the fixed point is attractive for decreasing $x$. Indeed, the full solution approaches the fixed point more and more closely at small $x\gg T/T_c$, as is clearly observed when
one solves the $T=0$ flow equations numerically. However, it is not entirely clear to us what \textit{physical} meaning should be ascribed to this kind of attractiveness.

Once the existence of the fixed point and its attraction for decreasing $x$ has been established, we can take $p^\star(z)$ as an initial condition for very small $x$ and forward integrate towards $x=1$. This last stage may be viewed as a renormalization flow of the field distribution to its low energy effective form, which attains an invariant form on intermediate time scales.

\subsection{Low $T$ solution}
At sufficiently low temperature, the fixed point controls the flow of $p(x,z)$ for $z=O(1)$ for all $x\gg T$, in the sense that one can take $p^\star(z)$ as an initial condition at $x\gg T$. In the limit $T\to 0$, one then solves Eqs.~(\ref{smeq1}-\ref{qpm1})
for all $x$ in an iterative manner analogous to the procedure described in the previous section.

At this stage, one may worry about the influence of boundary conditions for $p(x,z)$ at large $z$. However, it turns out that the fixed point is very stable and strongly attractive in the considered models, such that the precise
form of the boundary conditions has very little influence on the resulting flow. Their effect seems to be exponentially small in the value $z_{\rm bc}$ where the boundary conditions are imposed. A rigorous way to ensure small errors from boundary conditions at finite but low temperature is described in App.~\ref{app:numerics}. The latter allowed us to assess the good quality of the results we obtained with simple boundary conditions resulting from forward integrating the flow for $p$ in the regime $z\gg 1$ where $m={\rm sign}(z)$,
\begin{equation}
p(x,z)=z^{\alpha}
+\eta z^{\alpha-1}(2\lambda^\star-x \Delta h(x))+{\cal{O}}(z^{\alpha-2}).
\label{bcx}
\end{equation}
Here, $\Delta h(x)$ is given by
\begin{equation}
\Delta h(x)=2\int_0^x dx'\frac{\lambda(x')-\lambda^\star}{x'^{2}}.
\label{dhx}
\end{equation}

\subsection{Scaling of the density of states}
The fixed point distribution $p^\star(z)$ governs the low temperature behavior of the thermodynamic field distribution of metastable states, $p(x=1,z)=P(x=1,y=z\, T)/T^2$ in the sense that it leads to a universal scaling function $p(x=1,z)\ra \Psi_{\rm 3D}(z)$, which is shown in Fig.~\ref{f:universal_lowP}.
While the flow from the fixed point to $x=1$ (relating $p^\star$ and $\Psi_{\rm 3D}$) modifies the precise functional shape on the scale of $z\sim 1$, it preserves the large field asymptotics, see Fig.~\ref{f:pscaling}.
In particular, we may extract the asymptotic value for the prefactor of the parabolic pseudogap from the fixed point function at large $z$ where to very good approximation we have $p(x=1,z)\approx p^\star(z)= C z^2+O(z)$, with $C=0.02604$. This implies $P(y)=  C y^2+O(y)$ for $T\ll y\ll T_c$. Translating the local field to the energy to introduce additional particles (or to flip the local spin), $E\equiv 2y$, and restoring the lattice units $e^2/4\kappa \ell$ and $\ell$ for energy and length, respectively, we find for the single particle density of states
\bea
\label{asymptoticprefactorof ES3D}
\rho(E)\approx 8C\frac{\kappa^3}{e^6}E^2= 0.2083\frac{\kappa^3}{e^6}E^2,\quad T\ll E\ll T_c.
\eea
The coefficient of the parabolic term is significantly smaller than the value $3/\pi\approx 0.955$ predicted on the basis of the minimal stability requirement of a low energy configuration with respect to single particle hops~\cite{efros76,Baranovskii80}. This can be ascribed to the fact that our mean field theory takes into account the back reaction of many particles, which should indeed lead to a more stringent restriction for the local density of states. It is interesting to note that an estimate~\cite{MuellerIoffe04} based on a simple function $P(y)$ and imposing marginality, leads to $8C\approx 0.204$ very close to the exact value in Eq.~(\ref{asymptoticprefactorof ES3D})~\cite{fn6}.

\begin{figure}
\includegraphics[width=3.0in]{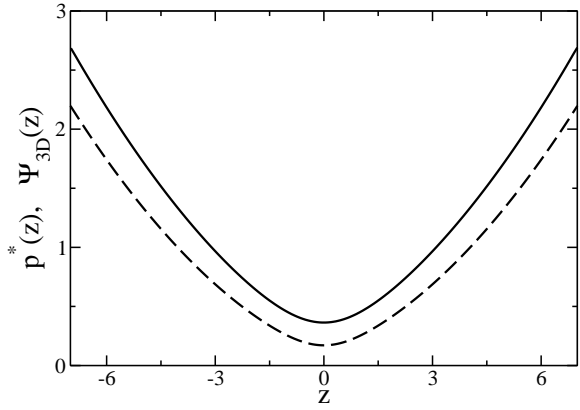}
\caption{Universal, asymptotic low temperature scaling function $\Psi_{\rm 3D}(z)= P(1,y=z\,T)/T^2=p(1,z)$ (full line) and the fixed point function $p^\star(z)$ (dashed line). Both have the same curvature at large $z$.}
\label{f:universal_lowP}
\end{figure}

\subsection{Corrections to scaling}
So far we have concentrated on the extreme low energy limit where we approximated $\psi(r_{\rm sc}^2)$ by the universal term in Eq.~(\ref{psi2}). In order for this term to dominate we need $r_{\rm sc}^2\gg (2C_{\rm latt})^2/\pi$. From the knowledge of the fixed point asymptotics $r_{\rm sc}^2 \approx (\beta x)^2 /\xi^\star$ for $x<x_P$, we estimate that we need to go to temperatures below $T\ll T^\star= \sqrt{\pi/\xi^\star}\, x_P/2|C_{\rm latt}|\approx 0.12 \equiv 0.03 e^2/\kappa\ell$  to ensure universal asymptotic fixed point behavior. At temperatures and energies of the order of or larger than $T^\star$, the non-universal, lattice dependent corrections (\ref{Nonuniversalcorrection}) influence the physical observables, in particular the density of states.

The effect of those corrections are clearly visible in the rather slow approach of our finite temperature data for $p(z,x=1)$ to a universal function $\Psi_{\rm 3D}(z)$, see the top panel of Fig.~\ref{f:Pyscaling}. This can be traced back to the deviation of $\psi_{\rm 3D}$ from its low energy asymptotics for $r_{\rm sc}\gg 1$. The convergence of the finite temperature data looks much more convincing if we repeat the above analysis, but approximate $\psi_{\rm 3D}(r_{\rm sc}^2)$ with an artificial power law $\tilde{A}r_{\rm sc}^{(2-2/{\tilde{\alpha}})}$ which fits the function best in the relevant regime of parameters $r_{\rm sc}$.
For the temperature range $30 \leq \beta \leq 100$, an exponent $\tilde{\alpha}\approx 2.34$ turns out to provide a reasonably good fit to $\psi_{\rm 3D}$, and to produce a much better scaling of $\tilde{p}(z)=\beta^{\tilde{\alpha}}P(y=z/\beta)$, as shown in the bottom panel of Fig.~\ref{f:Pyscaling}. As a consequence, it looks as if the density of states were governed by a larger exponent $\tilde{\alpha}$ at energies $E\geq T^\star$.

The apparent fixed point in the flow of average magnetizations and field distributions as a function of Sompolinsky time $x$ is well captured by the fixed point theory applied to the above artificial power law, as illustrated in Figs.~\ref{f:mscaling} and \ref{f:pscaling}. However, the genuinely universal low temperature fixed point predicted by Eqs.~(\ref{fpmeq},\ref{fppeq}) only applies to the very low energy and temperature regime.

A similar effective power law of the pseudogap was also found in numerical simulations of 3D Coulomb systems. The energies for which the density of states can be sampled reliably lies systematically above $T^\star$, because of finite size limitations. For this non-universal regime M{\"o}bius et al.~\cite{Moebius93} reported exponents $\tilde{\alpha}$ of the order of $2.2-2.4$ at the lowest reliable energies, in good agreement with our mean field prediction.

\begin{figure}
\includegraphics[width=3.0in]{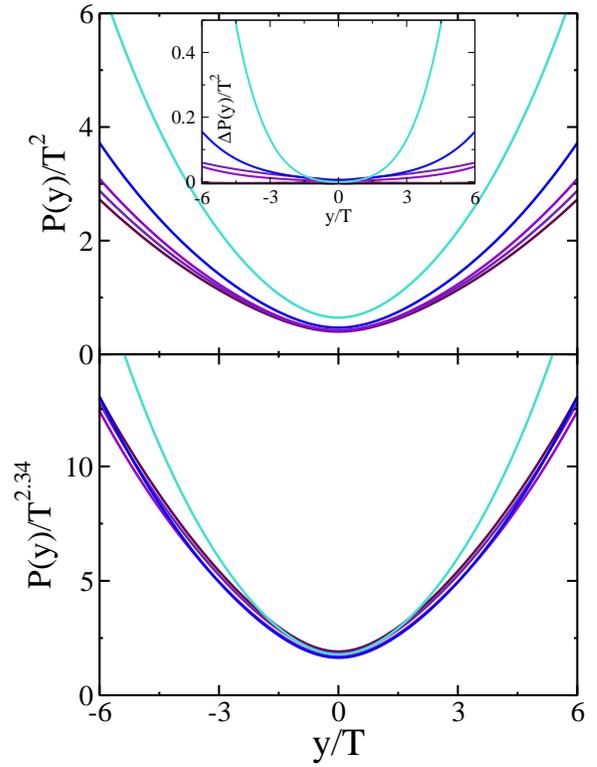}
\caption{(Color online) Scaled distribution of thermodynamic fields as a function of temperature. The curves correspond to $W=2$ and $\beta=20,40,60,80,100$ (from top to bottom).
Top panel: Scaling with the Efros-Shklovskii-exponent $\alpha=2$, $p(1,z)= P(1,y=z\,T)/T^2$. The inset demonstrates that the differences $\Delta P=|P_{W=2}-P_{W=10}|$ due to the strength of disorder tend to zero in the low $T$ limit, as expected from the universality induced by the fixed point.
Bottom panel: Improved scaling with an artificial power $\tilde{\alpha}=2.34$, $p(1,z)= P(1,y=z\,T)/T^{\tilde{\alpha}}$.}
\label{f:Pyscaling}
\end{figure}

\subsection{The plateau and the overlap distribution at $T=0$}

The $T\to0$ limit of the self-consistent coupling function $\lambda(x)$ is shown in Fig.~\ref{f:lambdax}. Starting at its fixed point value $\lambda^\star$ at small $x$,
it slightly decreases with increasing $x$, and finally vanishes rather abruptly at the plateau threshold $x_P=x_P(T=0) \approx 0.78$. The function $q(x)$ obviously exhibits an analogous behavior.

For $x>x_P$, $\lambda(x)=q(x)=0$ which implies that the overlap function $Q(x)$ has a plateau for $x_P<x<1$. This has interesting consequences:
According to Parisi's interpretation of the overlap function, $P(Q)=\left.[dQ/dx]^{-1}\right|_{Q(x)=Q}$ describes the distribution of overlaps among metastable states, if sampled according to their Boltzmann weight. The plateau implies that the lowest lying state (the ground state valley) takes a finite weight of the total Boltzmann weight, contributing a $\delta$-peak at the Edwards-Anderson overlap $Q_{\rm EA}$, corresponding to the self-overlap of typical low-lying states. Higher-lying states are separated from the lowest state at least by $O(T)$ in energy.
Similar features~\cite{fn7}
are known for the SK model~\cite{CrisantiRizzo02,Pankov06}.

\begin{figure}
\includegraphics[width=3.0in]{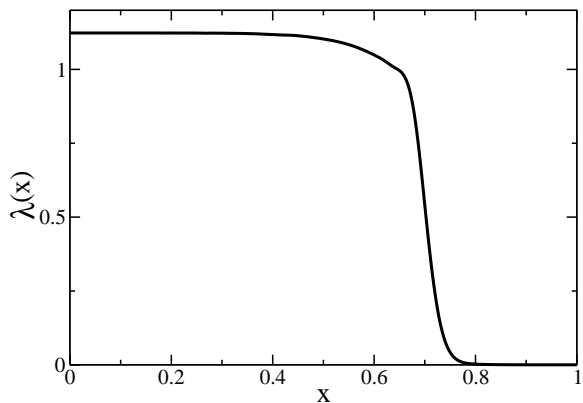}
\caption{$T=0$ solution of the coupling function $\lambda(x)=(\beta x)^{3}\dot \Lambda(x)/2\beta$ for the 3D electron glass. The fixed point controls the value $\lambda(x\ra 0)=\lambda^\star$. The drop to $0$ around $x_P\approx 0.78$ indicates a plateau in the coupling and overlap function.}
\label{f:lambdax}
\end{figure}

The scaling (\ref{qvsQdot}) suggests to introduce the
reduced phase space distance $\hat{d}=(1-Q)/T^3$, with the intra-state (Edwards-Anderson) distance $\hat{d}_{\rm EA}=0.8609$. The asymptotic fixed point $q(x)\ra q^\star$ implies a power law decay of the tails of the overlap distribution at low $T$,
\bea
P(\hat{d})&=&(1-x_P)\delta(\hat{d}-\hat{d}_{\rm EA})+x_P P_{\rm tail}(\hat{d}),
\eea
where $P_{\rm tail}(\hat{d})\sim \hat{d}^{-(2+\alpha)/(1+\alpha)}$. The low temperature limit of this distribution is plotted in Fig.~\ref{f:Pofd}.

\begin{figure}
\vspace{.0cm}
\includegraphics[width=3.0in]{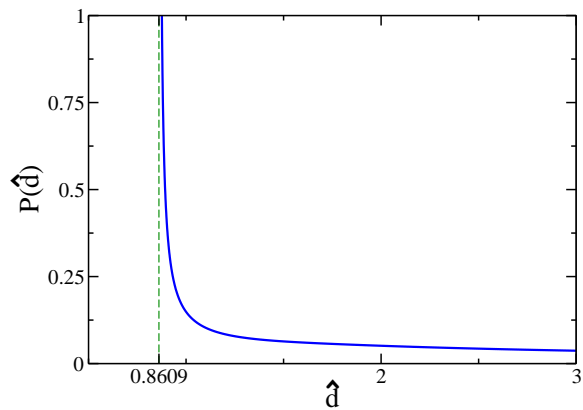}
\caption{(Color online) Equilibrium distribution of the rescaled overlap distance, $\hat{d}=(1-Q)/T^3$, in the 3D electron glass. There is a peak of strength $1-x_P$ at $\hat{d}_{\rm EA}=0.8609$ corresponding to the intrastate (Edwards-Anderson) overlap, and an asymptotic power law tail $P(\hat{d})\sim \hat{d}^{-(2+\alpha)/(1+\alpha)}=\hat{d}^{-4/3}$.}
\label{f:Pofd}
\end{figure}

\section{Physical interpretation of the asymptotic fixed point}
\label{s:FixedpointInterpretation}
Apart from providing accurate numerical values of low temperature characteristics, the asymptotic fixed point in the flow equations has several interesting physical consequences.

\subsection{Dynamical self-similarity}
\label{ss:dynSelfsimilarity}
Consider the distribution of local magnetizations averaged over a time scale $t_x$, $\rho(m;x)$. On intermediate time scales $x$, with $T/T_c\ll x<x_P$, the fixed point attracts the flow and we find to a good approximation
\bea
\label{invariantdistofm}
\rho(m;x)&\equiv& \int d y\, P(x,y)\delta(m-m(x,y))\nn\\
&\approx& \frac{1}{(\beta x)^3}\int d z\, {p^\star(z)}\delta(m-m^\star(z))\nn\\
&=&\frac{1}{(\beta x)^3}\rho^\star(m)\equiv T_{\rm eff}^3(x)\rho^\star(m),
\eea
with the invariant form factor
\bea
\rho^\star(m)=\left.\frac{p^\star(z)}{m^{\star\prime}(z)}\right|_{m=m^\star(z)}\,.
\eea
This applies for all $|1-m|$ such that $z=\left[m^\star\right]^{-1}(m)$ is not too large ($|1-m|$ not exponentially small), such that the fixed point function indeed applies.

Notice that the dynamics merely increases the number of active degrees of freedom (by the third power of the increasing factor $T_{\rm eff}/T$, as seems natural in presence of a parabolic pseudogap), without changing the overall shape of the distribution except in the tails. This result indicates an interesting self-similarity of the dynamics. We may actually speculate that after a suitable coarse-graining of time the dynamics will look essentially the same on all time scales, except that the actual number of active degrees of freedom gradually increases.

\subsection{Generalized fluctuation-dissipation relation}
\label{ss:FDT}
\subsubsection{Exact relation for global properties}
In equilibrium, the classical fluctuation-dissipation theorem (FDT) relates the integrated response function $\chi(t,t')$ and the correlation function $C(t,t')$ via
\bea
\label{FDTclassic}
\chi(t,t')= \beta\frac{\p C(t,t')}{\p t'}.
\eea
In equilibrium, time translation invariance holds and both functions only depend on $t-t'$. However, in a glass, time translation invariance is broken because the system remains out-of-equilibrium and "ages". That is, over longer and longer times it explores growing portions of phase space. In this case, the relation (\ref{FDTclassic}) no longer holds on large time scales, but is replaced by
\bea
\label{FDTCuKu}
R(t,t')=\beta x(C) \frac{\p C(t,t')}{\p t'}.
\eea
The extra factor $x(C)\leq 1$ only depends on the correlation $C$, provided both times are large, $t,t'\gg t_0$. The temperature in the FDT relation simply is replaced by $T_{\rm eff}(t,t')\equiv T/x(C(t,t'))$, which, at fixed initial time $t'$, may serve as a measure of the observation time $t$, as we used implicitly all along this paper.

The above relation (\ref{FDTCuKu}) actually \textit{defines} an "effective temperature" as the ratio $\frac{\p C(t,t')}{\p t'}/R(t,t')$.

Out of equilibrium dynamical behavior such as described above appears explicitly in the mean field solution of the Langevin dynamics of various spin glass models~\cite{CugliandoloKurchan93,CugliandoloKurchan94}.
For the case of glasses with continuous replica symmetry breaking, one can obtain the generalized FDT relation (\ref{FDTCuKu}) directly from Sompolinsky's anomalous response Ansatz~(\ref{anomalousresponse}).
In fact, (\ref{FDTCuKu}) is just a simple rewriting of the latter, if we note that $C(x)$ appearing in the FDT relation is the same as $Q(x)$ in Sompolinsky's dynamics. Therefore, for such mean field models, the $x$ in (\ref{FDTCuKu}) and the one appearing in the replica and Sompolinsky formalisms are in fact identical.

Even though these arguments may suffice to show that the generalization of the FDT relation is automatically built into Parisi's replica symmetry breaking scheme, it is useful to derive the relation directly from the flow equations. This will illustrate the meaning of the quantities $m(x,y)$ and $P(x,y)$ and the formalism introduced to solve the mean field equations.

In the spirit of Sompolinsky's approach, the integrated response function $\chi(t,t')=\int_{t'}^t R(t,t'') d t''$ takes the natural expression
\bea
\label{chitt}
\chi(t,t')\equiv \chi(x)=\int P(x,y)\frac{\p m(x,y)}{\p y}\, d y
\eea
where $x$ is associated to the pair of initial and observation times $(t',t)$ in the sense explained above. In the replica formalism one anticipates that this response is given by $\beta \langle s_as_b\rangle$.
By inspection of Eqs.~(\ref{SC1},\ref{SC2}), this suggests the relation
\bea
\label{chi=chi1}
\chi(x)=\chi_1(r_{\rm sc}(x)),
\eea
which we will confirm shortly.

The time averaged correlation function is similarly given by
\bea
\label{Ctt}
C(t,t')=\int P(x,y)m^2(x,y)\, d y \equiv Q(x).
\eea
Taking the time or $x$-derivative of both equations (\ref{chitt},\ref{Ctt}), one easily establishes the generalized FDT relation (\ref{FDTCuKu}) by using the flow equations and partial integration to show
\bea
\label{FDTfromParisi}
\frac{d} {d x}\chi_1(x) &=& \beta x \frac{d \Lambda}{d x} \int P(x,y) \left[m^\prime(x,y)\right]^2\, d y\nn\\
&=& -\beta x \frac{d Q}{d x}.
\eea
To prove the second relation, we have used Eq.~(\ref{qdot}).

Further, a comparison with Eqs.~(\ref{SCrew}) or (\ref{SClowT2}) establishes the equivalence conjectured in Eq.~(\ref{chi=chi1}).

\subsubsection{Local interpretation of $T_{\rm eff}$}

The above relation is an exact global relation which makes a statement about the correlations and response averaged over the whole system. On the other hand, the asymptotic fixed point allows us to establish a local, though only asymptotically exact, counterpart.

Consider the average {\em local} magnetization of a site $i$ seeing a frozen field $y_i$ over a time scale $t_x$ with the same restrictions on $x$ as earlier. In that regime, it is governed by the fixed point
\bea
\langle s_i \rangle_{t_x}\approx m^\star[y_i/T_{\rm eff}(x)].
\eea
Moreover, the time averaged response to an external field $h_{\rm ext}$ applied over a comparable time scale is governed by $T_{\rm eff}$ instead of $T$
\bea
\frac{\d \langle m_i \rangle _{t_x}}{\d h_{\rm ext}}=\frac{1}{T_{\rm eff}(x)} (m^\star)'[y/T_{\rm eff}(x)],
\eea
where $m^\star$ substitutes for the familiar $\tanh$ as a response function.

\section{Aging}
\label{s:Aging}
\subsection{Physical meaning of the plateau}
An appealing static interpretation of Parisi's Ansatz was found by M\'ezard et al.~\cite{MezardParisi85c}. They rederived the solution within a cavity approach, by assuming a hierarchical organization of states within clusters within even larger valleys etc. An essential ingredient is the distribution of the free energies of states relative to the cluster they belong to, the distribution of cluster energies relative to their valley, and so on. In order to recover Parisi's RSB scheme for the SK model in a $K$ step RSB approximation, one has to suppose  exponential distributions $\rho(F^{(i)})\propto \exp[x_{K+1-i}\, \beta F^{(i)}]$, where the $F^{(i)}$ are the free energies pertaining to a valley on the $i^{\rm th}$ level of the hierarchy ($i=1,\dots, K$). The parameters $1>x_K>\dots>x_1$ extremizing the free energy turn out to be identical to the optimal size of replica blocks in a $K$-step approximation. This gives a further interpretation of the variable $x$ as describing the energy landscape at a certain level of the hierarchy of valleys (which is explored by the system on the Sompolinsky time scale corresponding to $x$).

For experimental purposes, one would like to know how the immediate environment of a state is organized, since this is the local landscape being usually explored in glassy response measurements. According to the above and the ultrametric picture of phase space organization, the free energies of states close to a given metastable state are distributed exponentially, and parametrized by the largest parameter $x_K$ appearing in the above hierarchy.
This picture carries through to the case of continuous RSB where this largest value is given by the plateau threshold $x_P=\lim_{K\ra \infty} x_K$.

\subsection{The trap model and simple aging}
The information about the local neighborhood of a metastable state can be combined with Bouchaud's trap model~\cite{Bouchaud92,BouchaudDean95} in order to make a non-trivial prediction about the aging behavior of the electron glass.

The trap model assumes the dynamics of the glass to be a random trajectory among the set of states within one cluster of Parisi's hierarchy. The escape from a given state ("trap") $j$ is assumed to be Poisson distributed with an average escape time $\tau_j=\exp[\beta F_j]$, where $F_j$ is the free energy barrier associated with the $j^{\rm th}$ trap. Taking the above cavity interpretation of Parisi's ansatz as a motivation, the $F_j$ are  taken to be random variables with a probability density $\rho(F)\sim \exp[\beta x F_j]$, so far with a free parameter $x$. Finally, a real system's dynamics is expected to perform a natural ensemble average over such trap models, since different regions in space can be thought of as roughly independent trap models.

For an exponential barrier distribution, the trap model dynamics depends sensitively on the parameter $x$. For $x>1$ the system is ergodic, sampling all of traps in a time proportional to their total number. However, if $x<1$ aging effects occur, reflecting that the dynamics is weakly non-ergodic: While the system is not stuck in any given trap forever, it is nevertheless unable to explore the whole phase space in a finite time since the mean dwelling time in a trap diverges, $\langle\tau\rangle=\infty$.

To quantify these effects, one can look at a system that evolved from random initial conditions  for a time $t_w$. The probability to find the system still in the same trap at time $t$ as at $t_w$ has been calculated~\cite{BouchaudDean95} to decay asymptotically like
\bea
\label{decay}
C(t;t_w)\sim [t_w/t]^x,
\eea
where $t$ and $t_w$ are measured from the same origin of time.

The problem of connecting the phenomenological trap model with quantitative models for glasses lies usually in the lack of prediction for the aging exponent $x$.
However, from our low temperature solution of the electron glass problem, we obtain such a prediction as $x=x_P(T\ra 0)\approx 0.78$. Hereby, we think of the dynamics as exploring only the lowest level of hierarchy in an ultrametric organization of states. Moreover, we make the assumption that the states within a valley have a common threshold, as depicted in Fig.~\ref{f:trapmodel}. This allows us to identify the free energies $F^{(1)}$ of the cavity approach with the free energy barriers of an effective trap model, whose parameter $x$ corresponds to the plateau value $x_P$, the first (largest) $x$ for which there is a non-trivial (non-constant) structure of the Parisi matrix $Q(x)$.

\begin{figure}
\includegraphics[width=3.0in]{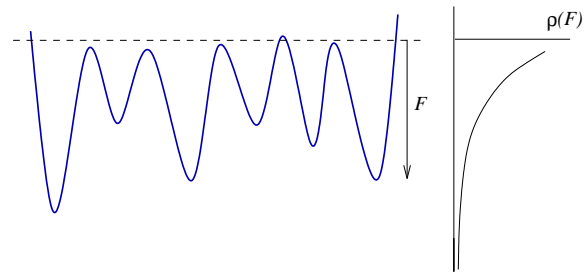}
\caption{(Color online) Schematic representation of the energy landscape assumed in the trap model. The depth of the traps is exponentially distributed, deeper traps being rarer than shallow ones. To make contact with the replica solution we have to assume that the states in a cluster have a common "ceiling" or threshold level beyond which the system easily escapes and becomes trapped in any other state within the cluster.}
\label{f:trapmodel}
\end{figure}

The above prediction may be relevant for aging experiments in electronic glasses. Indeed, recent measurements of electronic aging in 2D indium-oxide films~\cite{Ovadyahu06} found conductance relaxations which obey "simple aging", i.e., a scaling of the relaxation with $t/t_w$, with an asymptotic decay that can well be fitted to the power law (\ref{decay}) with exponents  $x=0.85-0.9$.

Similar experiments on granular aluminum~\cite{Grenet04} seem to be consistent with an exponent $x=1$. A possible reason may be the different nature of the Coulomb interactions in such a system, which is suggested by the unusual transport behavior in the hopping regime.~\cite{Grenet03}

\section{Discussion}
\label{s:Discussion}

\subsection{Experimental implications of the glassy state}
The glass transition should in principle be observable as a divergence of the non-linear capacitance. In practice, this will prove rather difficult to measure, even though attempt sin this direction were made~\cite{DonMonroe87,Monroe90}. It may be easier to look for other indications of the glass phase, such as out-of-equilibrium behavior or manifestations of the glassy criticality predicted for the whole low temperature phase. The criticality and marginal stability of typical metastable states arises from their approximate degeneracy with many other nearby valleys in the energy landscape. The latter leads to a large number of collective soft modes (low energy rearrangements) around local minima and saddles in configuration space.

To the extent that this prediction is not an artifact of the mean field approximation, it should have important physical consequences for the Coulomb glass. We expect the marginality to be reflected in the original lattice problem as a permanent criticality of the low temperature state, in the sense that the square of the connected two point correlator (\ref{squarecorrelator}) retains a power law decay even below $T_c$. This implies in particular that screening is not exponential. Instead the charge correlations decay slowly and even have a random signs at far enough distances. Similar behavior was indeed observed in numerical simulations~\cite{Baranovskii84} at $T=0$.

We infer the above scenario from a closely analogous situation in the SK spin glass, where the local approximation and Parisi's solution are known to be exact. In the same way as for the Coulomb glass, one can prove the marginal stability of the low temperature phase for all $T<T_c$ in the single site model. On the other hand, it is known from the analysis of the Thouless-Anderson-Palmer equations for all local magnetizations~\cite{ThoulessAnderson77} $m_i$, that the marginal stability of the single site approximation is directly related to an extensive number of gapless modes in the excitation spectrum around typical metastable minima~\cite{BrayMoore79}.
We conjecture that a similar connection holds between the marginality of the local approximation and the criticality of the lattice Coulomb glass.
If this is true, the presence of a large number of soft collective excitations will be a crucial ingredient to the understanding of glassy dynamics and relaxation. In the quantum version of the Coulomb glass, the marginality is closely related to the closure of the gap in the spectral function~\cite{dalidovich02}, and one expects the presence of the associated collective soft modes to have an important impact on conductivity and DC variable range hopping in particular.

\subsection{Extension to 2D electron glasses}
In two-dimensional electron glasses with $1/r$ interactions, the mean field approximation is less well justified than in 3D due to the reduced number of effective neighbors. If we nevertheless carry out analogous approximations as in 3D, we obtain a similar single site problem, though with a function $\psi_{\rm 2D}[r_{\rm sc}^2]\approx 2\pi\ln{[r_{\rm sc}(x)^2]}$. Because of the logarithmic corrections, there is no genuine fixed point in the flow equations as in 3D.
However, in order to qualitatively describe the distribution of fields around a given $T\ll \overline{y} \ll 1$, we can approximate $\psi_{\rm 2D}\approx {\rm const.} \approx A \approx 2\pi\ln(1/\overline{y})$, which corresponds to an exponent $\alpha=1$, reducing the single site model to an effective SK model with random fields. From the solution of the latter~\cite{SommersDupont84,Pankov06} we know that $p^\star(z)= 0.30 |z|/A$, and we thus expect the 2D electron glass to exhibit a quasilinear pseudogap $P(y)\approx 0.3 |y|/2\pi \ln{(1/|y|)}$ at low energies, with logarithmic corrections suppressing small fields.

\subsection{Limitation of the mean field approximation}

In the mean field approach we neglect certain correlations among the neighbors of a given "cavity" site. More precisely, the response of neighboring sites to the presence of a cavity spin is treated as essentially independent, the only correlations retained being ring diagrams. This approximation does not include degrees of freedom such as electronic dipoles~\cite{Baranovskii80,Efros00}. These are pairs of sites, one occupied the other empty, with individual local fields $|h_{1,2}|\gg T$, but with an excitation energy $h_1+h_2-{\cal J}_{ij}\lesssim T$ for swapping the particle hole pair. Such dipoles can be defined unambiguously only at temperatures sufficiently below $T_c$.

The interaction of single (active) sites with such degrees of freedom is not, or not entirely contained in the single site approximation. However, it is possible that the behavior of single sites is actually dominated by the interaction with other active sites, rather than by the interaction with dipoles, which is assumed in our mean field approach. Indeed, we obtain very reasonable results for the single particle density of states. However, a clear indication for the neglect of dipolar degrees of freedom in the mean field theory is found in the specific heat.
While the single site model would predict $C_V\sim T^D$, it is known that dipoles dominate the specific heat of the full lattice model, with an almost linear temperature dependence~\cite{ESeedisorder,mobius96,Efros00}.

It is a very difficult open problem to build a self-consistent mean field theory which would include such non-local objects. They must be expected to be dynamically generated as the system settles into low energy metastable states, their location and orientation depending on the specific metastable state under consideration, which makes the problem very difficult to tackle.

\subsection{Comparison with numerical studies}
\label{ss:compwithnumerics}

\subsubsection{Glass transition}
So far numerical studies for the model~(\ref{Hamiltonian}) with on-site disorder have remained inconclusive as to the existence of a glass transition, even though a transition was claimed to exist in a 3D system with positional disorder but no on-site randomness~\cite{GrannanYu93}. We hope that future studies searching for a divergent non-linear capacitance and/or critical spin-spin correlation functions will eventually settle the question about the existence of a thermodynamic glass transition in the 3D lattice model.

\subsubsection{Density of states}
The local density of states and the Efros-Shklovskii pseudogap has been subject of  rather extensive studies in the numerical literature~\cite{Moebius93,sarvestani95}.
It has already been shown in Ref.~\onlinecite{pankov05} that the temperature-dependent distribution of instantaneous fields, $P(h)$, reproduces remarkably well the numerical results in the high temperature phase of the moderately disordered model studied in Ref.~\onlinecite{sarvestani95}. Here we briefly discuss the low temperature and low energy results.

As we have argued above the fixed point prediction~(\ref{asymptoticprefactorof ES3D}) for asymptotically low temperatures applies to very low energies which are far from reach in numerical simulations. However, our results obtained at intermediate energies $0.01<E<1$ compare reasonably well with the numerical data of Ref.~\onlinecite{Moebius93}, cf.,~Fig.~\ref{f:mobiuscmp}. One should however bear in mind that the latter cover only small to moderate disorder strengths for which our mean field theory is not easy to justify.
Nevertheless, as mentioned before, the apparent larger pseudogap power $\alpha\approx 2.2-2.4$ in the numerical data is well reproduced by mean field theory as a consequence of lattice effects.

While our theory predicts an essentially disorder independent density of states already at moderately low energies, the one found in the simulations increases with bare disorder. This might be either due to the increasing difficulty to anneal a sample with stronger disorder in a simulation, or it could be a further effect of the lattice discreteness which affects the non-universal regime of energies above $T^\star$.

The theoretical prediction extracted from our lowest temperature data for the density of states, $P(h;T=0.01)$, lies somewhat below the numerical data. This is likely to be ascribed to the clustering of sites with energies close to the chemical potential, which was observed long ago in simulations of Davies et al.~\cite{Davieslee84}.
This effect is not captured correctly in our mean field description (and it is also neglected in the ES stability argument). Due to the clustering, the numerical data suggests a greater number of active sites than there really are, e.g., for the purpose of hopping conductivity. Indeed, it is not possible to fill two sites in the same small cluster of empty low energy sites without paying a lot of energy. We believe that in the mean field theory such small clusters effectively occur as a single soft site which leads to a lower $P(h)$ than seen numerically, while the overall behavior as a function of energy is well reproduced.

\begin{figure}
\includegraphics[width=2.8in]{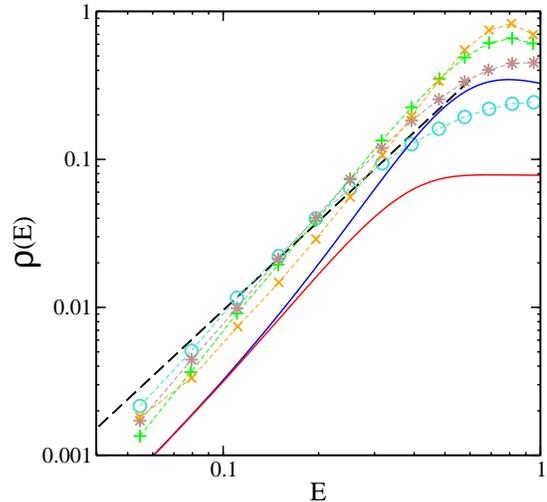}
\vspace{.2cm}
\caption{(Color online) Symbol lines: 3D data taken from Ref.~\onlinecite{Moebius93}, corresponding to uniformly distributed disorder with variance $\tilde{W}=\langle (\epsilon/2)^2\rangle = 0.29, 0.58, 1.15, 2.30$ in units of $e^2/4\kappa\ell$ (from bottom to top at low energies). The full lines (blue and red) are our mean field predictions for $P(h)$ at $T=0.01$, for Gaussian disorder with $W=2$ and $W=10$, respectively. Dashed line: Prediction from the ES stability argument~\cite{Baranovskii80}, $\rho(E)=(3/\pi) E^2$. In the plot, we use the standard units $e^2/\kappa\ell$ and $\ell$ for energy and length, respectively.}
\label{f:mobiuscmp}
\end{figure}

However, such a clustering phenomenon should not affect the {\em ratio} of the density of states at different energies and temperatures. In the literature, the ratio $r\equiv \rho(E=0,T=t)/\rho(E=t,T=0)$ was studied numerically for small $t\gtrsim 0.05$ (in energy units of $e^2/\kappa\ell$), and a rather large value $r\approx 11$ was obtained~\cite{LevinNguen87} by extrapolation to $t=0$. While this ratio could not be reproduced by finite $T$ arguments along Efros' self-consistency argument~\cite{MogilyanskiiRaikh89}, the present mean field theory comes much closer to the numerical finding [with our notation and units one has $r=P(h=0,T=4t)/P(h=2t,T=0)$]. In particular our finite temperature results predict the ratios $r(t=0.018)\approx 10.8$ and $r(t=0.0125)\approx 8.5$, which eventually decreases towards the extreme low energy value $r(t \ra 0)\approx 7.3$ which can be extracted from the  fixed point theory.

\subsubsection{Dynamics}
It would be interesting to investigate the dynamic aspects of the density of states, or rather, of the distribution of time-averaged site magnetizations $m=\tanh(\beta y)$, in order to check for the dynamical self-similarity predicted by the replica theory and its interpretation {\`a} la Sompolinsky. The same applies to the aging dynamics in the SK model, where the prediction of dynamical self-similarity might serve as a new check for the validity of the dynamic interpretation of the mean field solution.

\section{Conclusion}
\label{s:conclusion}
We have presented a detailed study of the mean field theory for electron glasses in 3D. A glass transition is predicted once the temperature becomes low enough for fluctuations in the Thomas-Fermi screening to become critically large. The low temperature phase is expected to be glassy and remain critical throughout due to a permanent marginal stability of typical metastable states. The latter also assures the saturation of Efros-Shklovskii's bound for the single particle density of states.

Solving the mean field equations in the low temperature phase and in the asymptotic $T\ra 0$ limit, we have obtained new quantitative predictions for the evolution of the density of states and the compressibility. We have given a physical interpretation of the asymptotic fixed point occurring in Parisi's flow equations in terms of a dynamical self-similarity of the distribution of time averaged magnetizations.

The low $T$ solution exhibits a plateau threshold $x_P$ in the overlap function $Q(x)$, which we have argued to describe the energy landscape in the vicinity of a given metastable state. Assuming a dynamics like in the trap model, we have used this information to predict the asymptotics of the aging function after an excitation.

This mean field approach is in principle amenable to include quantum tunneling between localized states~\cite{dalidovich02,dobrosavljevic03}. The resulting quantum glass will be an interesting candidate to study effects of quantum glassiness on transport and slow dynamics.

\vspace{10pt}
We acknowledge discussions with V.~Dobrosavljevi{\'c}, A.~L.~Efros, L.~B.~Ioffe, M.~M{\'ezard}, Z.~Ovadyahu, D. Sherrington, and B.~I.~ Shklovskii. This work was in part supported by grant PA002-113151 of the Swiss National Science Foundation.

%
%

\appendix

\section{Mapping to single site model}
\label{app:maptoSS}

In this appendix we show how to derive an effective local theory
using a cumulant expansion.

Let us consider a Hamiltonian $H(\phi)$ describing a generic problem
with two body interactions. The fields $\phi$ represent the available
degrees of freedom (Ising spins in our case).
We separate the Hamiltonian into a local part, $H_1(\phi)$, and
a non-local part $H_2(\phi)$ describing the two body interactions
\begin{equation}
H_2(\phi)=\frac{1}{2}\sum_{r,r'}\phi(r){\cal J}(r,r')\phi(r'). \label{s1}
\end{equation}
The variable $r$ is usually a space index, but can also incorporate other
coordinates such as a replica label.

We now perform a cumulant expansion of the partition function in terms of the nonlocal term $H_2(\phi)$, expanding around the local
limit described by $H_1(\phi)$. In a graphic representation,
each diagrams consists of lines representing the nonlocal matrix ${\cal J}(r,r')$ and vertices. An $n$-point
vertex stands for a bare cumulant $M^0_n$, defined as
\begin{equation}
M_n^0=\left.\frac{d^n}{dh^n}\right|_{h=0}\ln{\langle \exp{(h\phi)}\rangle_{H_1}}.
\label{mbare}
\end{equation}
The diagrammatic expansion is organized in a transparent way by introducing the "one-particle irreducible" $n$-point vertex $M_n(r_1,...,r_n)$ where the $r_i$ denote the space (and possibly further internal) indices at which the $n$ external legs are attached. The two-point correlator can then be expressed conveniently as a geometric series,
\begin{equation}
\label{propagator}
G(r,r')=\langle\phi(r)\phi(r')\rangle
=\left(M_2^{-1}+\beta{\cal J}\right)^{-1}_{rr'}.
\end{equation}

The effective local theory used in the paper can be justified and derived
in the limit of large disorder, or alternatively by invoking a large space dimensionality and performing an expansion in $1/D$.
Analyzing the structure of the leading irreducible vertex contributions, one finds that their $k$-dependence is negligible on the scales of interest, i.e., $k\lesssim T_c$. This suggests to treat all irreducible vertices as local,
$M_n(r_1,...,r_n)=0$ unless all $r_i$ refer to the same site. In particular, for the local correlator Eq.~(\ref{propagator}) simplifies to
\bea
\label{Gloc}
G=G(r,r)=\frac{1}{{\cal N}_q}\sum_q\left(M_2^{-1}+\beta{\cal J}_q\right)^{-1},
\eea
where we performed a decomposition into the ${\cal N}_q$ eigenmodes $q$ of the matrix ${\cal J}(rr')$.
The correlation function can simply be written as $G=M_2-M_2\Delta M_2$, defining the dressed bond operator
\bea
\Delta=\frac{1}{{\cal N}_q}\sum_q\left(M_2+(\beta{\cal J}_q)^{-1}\right)^{-1}.
\eea
By restricting to local subdiagrams, the irreducible vertices can be expanded into a perturbation series in terms of the local dressed bond $\Delta$ and the bare cumulants $\{M_n^0\}$, $M_n={\cal M}_n[\Delta,\{M_n^0\}]$.

Repeating a similar analysis for a local model with Hamiltonian
$H^{({\rm loc})}(\phi)=H_1(\phi)+H_2^{({\rm loc})}(\phi)$, we find expressions very similar to the above ones, with spatial indices suppressed.
In particular, for a two body interaction $H_2^{({\rm loc})}=\phi\Lambda^{({\rm loc})}\phi/2$, we find
$M_n^{({\rm loc})}={\cal M}_n[\Delta^{({\rm loc})},\{M_n^0\}]$ and
$\Delta^{({\rm loc})}=\left[M_2^{({\rm loc})}+
\left[\beta\Lambda^{({\rm loc})}\right]^{-1}\right]^{-1}$, with the same
bare cumulants $\{M_n^0\}$ and the same functionals ${\cal M}_n$ as for the lattice model.

From these observations it becomes clear
that the lattice correlation functions can be computed with the help of the local model on which we impose the self-consistency condition
$\Delta(r,r)=\Delta^{({\rm loc})}$, or equivalently, $M_2(r,r)=M_2^{({\rm loc})}$.
From Eq.~(\ref{Gloc}) and its analog in the
local model, $[G^{({\rm loc})}]^{-1}=[M_2^{({\rm loc})}]^{-1}+\beta \Lambda^{({\rm loc})}$, we can rewrite the self-consistency equation as
\begin{equation}
G=\sum_q\left(G^{-1}-\beta\Lambda^{({\rm loc})}+\beta{\cal J}_q\right)^{-1},
\label{sceq}
\end{equation}
where $G=G(r,r)=G^{({\rm loc})}$.

In the main part of the paper we have adopted the notations:
\bea
G&\to& Q, \\
\Lambda^{({\rm loc})}&\to& -\beta\Lambda,\\
H_1&\to& -\frac{1}{2}\beta W^2\sum_{a,b}s_a s_b,\\
H_2&\to& \sum_{i>j,a,b}s_i^a\mathcal{J}_{ij}s_j^b,\\
H_2^{({\rm loc})}&\to&-\frac{1}{2}\beta\sum_{a,b}s^a \Lambda^{ab}s^b.
\eea
Further, the irreducible two-point vertex is identified as the irreducible polarizability divided by $\beta$,
\bea
\label{M2Pi}
M_{2,ab}\to \Pi_{ab}/\beta.
\eea
In Ref.~\onlinecite{MuellerIoffe04} the notation ${\bf \Pi}/\beta \equiv (\beta^2 W^2{\bf {\cal I}}+{\bf \Sigma})^{-1}$ was used. This was suggestive to indicate two contributions to the "self-energy" $\beta/{\bf \Pi}$ occurring in the field theory for fields conjugate to the spins: a bare quadratic term $\beta^2 W^2{\bf {\cal I}}$ and an additional  dressing self-energy ${\bf \Sigma}$. The latter can conveniently be shown to be predominantly local for large $W$.

\section{Computation of $J_k$ on a hypercubic lattice}
\label{app:LatticeFT}

In order to compute the Fourier transform of $1/r_{ij}$ on a hypercubic lattice, we use Ewald type summation
\begin{equation}
J_k=\sum_{x_l\ne 0}\frac{e^{ikx_l}}{|x_l|}
=\sum_{x_l\ne 0}I(0,+\infty;x_l)e^{ikx_l},
\label{jk}
\end{equation}
where
\begin{equation}
I(t_1,t_2;x)=\frac{1}{\sqrt{\pi}}
\int_{t_1}^{t_2} t^{-\frac{1}{2}}e^{-tx^2}dt.
\label{it1t2}
\end{equation}
To obtain well behaved integrals we have to split $I(0,+\infty;x)$ into
two terms $I(0,\epsilon;x)$ and $I(\epsilon,+\infty;x)$ and switch
to a momentum sum in the first term. To this end we use the representation
\begin{equation}
\int d^dx\sum_{x_l\ne 0}\delta(x-x_l)
=\int d^dx \sum_{G_l}\left[e^{iG_lx}-\delta(x)\right],
\label{switch}
\end{equation}
where in the right hand side the summation is over the vectors of the reciprocal lattice.
We then integrate out $x$ and change the variable of integration in $I$ to
$t\to 1/t$. In this way we find
\bea
J_k&=&\left(\frac{\pi}{\epsilon}\right)^{\frac{d-1}{2}}
\sum_{G_l}\phi_{\frac{d-3}{2}}\left[\frac{(G_l+k)^2}{4\epsilon}\right]-2\nn\\
&&+\left(\frac{\pi}{\epsilon}\right)^{\frac{-1}{2}}
\sum_{x_l\ne 0}\phi_{-\frac{1}{2}}(\epsilon |x_l|^2)e^{ikx_l},
\label{jk1}
\eea
where $\phi_n$ are Misra functions, defined as
\begin{equation}
\phi_n(z)=
\int_{1}^{+\infty} t^{n}e^{-zt}dt.
\label{misrafunc}
\end{equation}
Note that for a hypercubic lattice with lattice spacing $1$, one has
$G_l=2\pi x_l$. Setting $\epsilon=\pi$, we can rewrite
Eq.~(\ref{jk1}) as
\bea
\label{jk2}
&&J_k=\phi_{\frac{d-3}{2}}
\left[\pi\left(\frac{k}{2\pi}\right)^2\right]-2\\
&&\quad+\sum_{x_l\ne0}\left[\phi_{-\frac{1}{2}}
\left(\pi |x_l|^2\right)e^{ikx_l}
+\phi_{\frac{d-3}{2}}
\left(\pi\left|x_l+\frac{k}{2\pi}\right|^2\right)\right].\nn
\eea
In the limit $k\to 0$ the first term is dominant and
yields $4\pi/k^2$ in $D=3$ dimensions.

\section{SK model with random fields}
\label{app:SK}

The Sherrington Kirkpatrick model is the mean field version of spin glasses, describing $N$ spins, all coupled mutually with random Gaussian interactions $J_{ij}$ with zero mean and variance $\langle J_{ij}^2\rangle=J^2/N$,
\bea
\label{SKRF}
H=\frac{1}{2}\sum_{ij}s_iJ_{ij}s_j+\sum_i s_i h_i.
\eea
We also include Gaussian random fields $h_i$ with variance $\langle h_{i}^2\rangle=W^2$, to obtain a model that closely resembles the electron glass problem.

For the SK model the locator approximation~\cite{BrayMoore79} is exact in the large $N$ limit.
The irreducible polarizability can easily be shown to be diagonal to leading order in $N$. This allows for the same local approximation as we have used in the electron glass problem, the only difference being that the interaction matrix $J_{ij}$ has a different eigenvalue spectrum,
\bea
\rho(\lambda)\,d\lambda=\frac{2}{\pi J}\sqrt{J^2-\lambda^2}\, \Theta(J-|\lambda|)\,d\lambda,
\eea
which is normalized such that $\int \rho(\lambda)\, d\lambda=1$.

This replaces the "mode density" of the electron glass
\bea
\label{rhoSK}
\rho(\lambda)\,d\lambda=\int \frac{d^Dk}{(2\pi)^D} \delta(\lambda-J_k)\, d\lambda.
\eea
By mapping to a single site model of the form
\bea
\beta H_0= -\frac{\beta^2}{2}\sum_{ab}s^a(\Lambda_{ab}+W^2)s^b,
\eea
one obtains the self-consistency condition for the SK model with random fields,
\bea
\label{SCSK}
Q_{ab}=\langle s_a s_b\rangle_{H_0}&=&\int \left.\frac{d\lambda\, \rho(\lambda)}{\beta \lambda-\beta/{\bf \Pi}}\right|_{ab}\nn\\
&=&
[-\beta^2 {\bf \Lambda} -\beta/{\bf \Pi}]^{-1}_{ab}.
\eea
In this case, the integral over $\lambda$ can be done explicitly. After a little algebra one recovers the saddle point equation from the standard replica treatment
which imposes that the overlap matrix and the effective coupling matrix coincide, $Q_{ab}=\Lambda_{ab}$.

The further steps to solve (\ref{SCSK}) are analogous to the case of the electron glass. In particular, the glass transition is given by the solution of
\begin{eqnarray}
\label{ATlineSK}
\beta^2 \int_{-\infty }^{\infty }dy \frac{P(y; Q_{\rm RS})}{\cosh^4 \left( \beta y\right)}  =1,
\end{eqnarray}
with
\bea
Q_{\rm RS}&=&\int\d y \,P_{{\rm RS}}(y; Q_{\rm RS}) \tanh^2(\beta y),\\
P_{\rm RS}(y; Q_{\rm RS})&=&\frac{\exp[-y^2/2(W^2+Q_{\rm RS})]}{\sqrt{2 \pi (W^2+Q_{\rm RS})}}.
\eea
From these equations it is easy to establish the scaling of the Almeida Thouless-line $T_c\sim 1/W$ at large disorder.

In Fig.~\ref{f:SKfigs} we illustrate the universality (disorder-independence) of the linear pseudogap, and its scaling behavior $P(y)= T\Psi_{\rm SK}(y/T)$ at low temperature.
\begin{figure}
\includegraphics[width=3.0in]{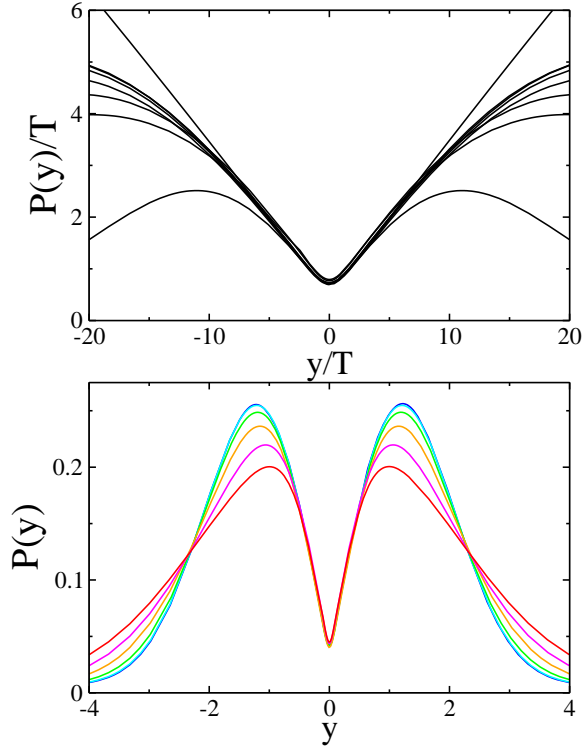}
\vspace{.2cm}
\caption{(Color online) Universality of the SK model in random fields. Top: Scaling of the local field distribution $P(y)$ for various temperatures $10\leq \beta\leq 60$ and disorders $0\leq W\leq 1.5$.
Bottom: Unscaled distribution $P(y)$ at $\beta=20$ for various disorder strengths, $W= 0,0.3,0.6,0.9,1.2$.}
\label{f:SKfigs}
\end{figure}

\section{Derivation of the Onsager term}
\label{app:Onsagerterm}
The Onsager term $h_O$ represents the response of the environment to a spin at a given site $0$. The latter should thus be excluded from all ring diagrams in (\ref{generalizedOnsager2}).
In order to establish a relation between $h_O$ and the self-consistent coupling matrix $\Lambda_{ab}$ we analyze the diagrammatic expansion for the local overlap
\bea
\label{sumforQ}
\beta Q^{ii}_{ab}&=&\left[\Pi- \Pi{\cal J}\Pi +\Pi {\cal J}\Pi{\cal J}\Pi-+\dots\right]^{ii}_{ab}
\eea
where the matrix products in the brackets involve sums over replicas and {\em all} sites, including $i$. $\Pi_{ab}\equiv \beta M_{2,ab}$ is the irreducible local polarizability or two point vertex. The result is independent of the site $i$ due to the disorder average.

Let us now define the "Onsager matrix"
\bea
\tilde{h}_{O,ab}&=&\left[{\cal J}\Pi{\cal J}
-{\cal J}\Pi{\cal J}\Pi{\cal J}+-\dots\right]^{ii}_{ab},
\eea
where all internal polarizabilities are restricted to sites $j\neq i$. With this matrix we can rearrange the sum (\ref{sumforQ}) according to the number of intermediate visits of the site~$i$,
\bea
\beta Q^{ii}_{ab}&=& \left[\Pi + \Pi \tilde{h}_O \Pi+ \Pi \tilde{h}_O \Pi \tilde{h}_O \Pi+...\right]_{ab} \nn\\
&=&\left[\frac{1}{\Pi^{-1}- \tilde{h}_O}\right]_{ab}.
\eea
Comparing to Eq.~(\ref{SP}) we see that in the mean field approximation, the Onsager matrix is essentially just the coupling matrix, $\tilde{h}_{O,ab}=\beta\Lambda_{ab}$.

In order to obtain the (average) Onsager back reaction in a given metastable state, we observe that the local, connected spin correlator calculated within a single state is given by $Q_{aa}-Q_{a\ra b}\equiv \tilde{Q}-Q(x=1)=[Q](x=1)$. The associated Onsager term is therefore
\bea
h_O&=&\tilde{h}_{O,aa}-\tilde{h}_{O,a\ra b}=\beta[\tilde{\Lambda}-\Lambda(1)]\nn\\
&=&\chi_1^{-1}(\kappa_C)-\frac{1}{\kappa_C}\approx 2\sqrt{\pi \kappa_C},
\eea
where we have used Eqs.~(\ref{SCrew},\ref{rsc}) in the second line and the asymptotic expression (\ref{fasym}) for the last step.

Note that for the SK model where $\Lambda\equiv Q$ one correctly recovers $h_O=\beta[\tilde{Q}-Q(1)]=\beta(1-Q_{\rm EA})$.

\section{Replica Fourier transform}
\label{app:RFT}

The replica Fourier transform~\cite{ReplicaFT1,ReplicaFT2} performs the diagonalization of a generic Parisi matrix ${\cal Q}$ to a representation in which it is straightforward to compute any analytic function of the matrix.

An ultrametric matrix with $K$ steps of replica symmetry breaking is fully described by its diagonal entries $Q_{aa}=\tilde{Q}$ and off-diagonal blocks with entries $Q_K,Q_{K-1},\dots,Q_1,Q_0$ in order of their distance from the diagonal. Such a matrix can be written as
\begin{equation}
\label{standard}
{\cal Q}=Q_0{\cal R}_0 +\sum_{k=1}^K (Q_k-Q_{k-1}) {\cal R}_{m_k} +(\tilde{Q}-Q_{m_K}) \,\rm{id},
\end{equation}
where $0<m_1<\dots<m_K<m_{K+1}\equiv 1$ are the sizes of the replica clusters in the hierarchy, and ${\cal R}_m$ denotes a $n\times n$ matrix with $n/m$ blocks on the diagonal each of which has all entries equal to one.

To compute functions of such a matrix, it is convenient to introduce the replica Fourier transform
\begin{equation}
\label{RFTform}
{\cal Q}=Q_0{\cal R}_0 -\sum_{k=1}^K \left[Q\right]_k {\cal A}_{k} +Q_c \,\rm{id},
\end{equation}
with the matrices ${\cal A}_k\equiv {\cal R}_{m_{k+1}}/m_{k+1}-{\cal R}_{m_k}/m_k$ for $k=1,\dots, K$, which satisfy orthonormality relations
\begin{equation}
{\cal A}_i\cdot {\cal A}_j=\delta_{ij}{\cal A}_i.
\end{equation}
This makes it is easy to compute any analytic function ${\cal G}=f({\cal Q})$ by Taylor expansion.

In the representation (\ref{RFTform}) one easily finds
\bea
\label{discFourier}
{\cal G}&=& f^\prime(Q_c)\,Q_0{\cal R}_0 \\
&&+\sum_{k=1}^K \left[f(Q_c-\left[Q\right]_k)-f(Q_c)\right]\, {\cal A}_{k} +f(Q_c) \,\rm{id}.\nn
\eea

The relation between $\left[Q\right]_k$ and $Q_k$ is
\begin{eqnarray}
Q_l&=&Q_0+\sum_{k=1}^{l-1} \left[Q\right]_k\left(\frac{1}{m_{k}}-\frac{1}{m_{k+1}}\right)+\frac{\left[Q\right]_l}{m_l}, \\
\tilde{Q}&=&Q_0+\sum_{k=1}^{K} \left[Q\right]_k\left(\frac{1}{m_{k}}-\frac{1}{m_{k+1}}\right)+Q_c,
\end{eqnarray}
implying
\begin{eqnarray}
\label{discretedqds}
Q_{l+1}-Q_l&=&\frac{\left[Q\right]_{l+1}-\left[Q\right]_{l}}{m_{l+1}}, \\
\tilde{Q}-Q_K&=&Q_c- \left[Q\right]_K.
\end{eqnarray}

In the continuum limit ($K\rightarrow \infty$, $Q_k\rightarrow Q(x=m_k)$)
Eqs.~(\ref{discretedqds}) turn into
\begin{equation}
\label{dqdx}
\frac{d[Q]}{dx}=x\frac{dQ}{dx}.
\end{equation}
Integrating, we find the continuous replica Fourier transform
\begin{eqnarray}
\label{qtos1}
\left[Q\right](x)&=&x Q(x)-\int_{0}^x Q(y) dy,\\
\label{qtos2}
Q_c&=&
\tilde{Q}-\int_0^1 Q(x) dx.
\end{eqnarray}

The continuous Fourier components of ${\cal G}=f({\cal Q})$ can be read off from (\ref{discFourier}) as
\begin{eqnarray}
\label{RFT1}
G_c&=&f(Q_c),\\
\label{RFT21}
\left[G \right](x)&=&f(Q_c)-f(Q_c-\left[Q\right](x)),\\
\label{RFT3}
G_0&=&f'(Q_c)Q_0,
\end{eqnarray}
which implies the useful identity
\begin{eqnarray}
\label{RFT2}
G_c-\left[G \right](x)&=&f(Q_c-\left[Q \right](x)),
\label{identity}
\end{eqnarray}
which is used repeatedly in the main part of the paper.

\section{Numerical Implementation of Parisi's equations}
\label{app:numerics}

In this appendix we describe several details of the numerical integration
of the flow equations~(\ref{mdot},\ref{Pdot}), which are subtle in particular in the low temperature limit. The integration of the flow equations forms the crucial part in the iterative scheme outlined in the main text.

Given $\Lambda(x)$ we first perform the integration of the flow for $m(x,y)$. Since the significant features of $m$ live on the scale $y\sim1/(\beta x)$, it is convenient to transform to the new variable $z=f(\beta x) y$ with $f(b)\sim b$ for large $b$. We have used $f(b)= b+c$ with $c$ a constant of order $10$, which yields a smooth behavior of $m(x,z)$, also at low $x$.

In a second step we integrate the flow of the distribution function $P(x,y)$. This is tricky at low temperature because there are at least two disparate energy scales, $e^2/\kappa\ell$ and $T$, which are both characteristic scales for $P(y)$. If the disorder $W$ is strong, it introduces an even larger scale, on which $P(y)$ varies significantly.

This problem can be solved by noting that to high accuracy we can set $m(x,y)={\rm sgn}(y)$ when $f(\beta x)|y|\gg 1$, where $m$ approaches $\pm 1$ exponentially. In this region of fields the flow equation~(\ref{Pdot}) simplifies to the linear partial differential equation
 \begin{eqnarray}
\label{Pdotsimple}
\dot{P}(y;x)&=&\dot{\Lambda}(x)\left[\frac{P''(y;x)}{2}-s\, \beta x P'(y;x)\right],
\end{eqnarray}
where $s\equiv {\rm sgn}(y)$. Its kernel is
\bea
\label{kernel}
G(x,y;x',y')=\frac{\exp\left[-\frac{(y-y'-s\beta\int_{x'}^x u\dot{\Lambda}(u)d u)^2}{2[\Lambda(x)-\Lambda(x')]}\right]}{\sqrt{2\pi[\Lambda(x)-\Lambda(x')]}},
\eea
with the help of which we can propagate initial conditions from $P(x=0,y)$ to finite $x$ in order to obtain boundary conditions for large $y\gg 1/f(\beta x)$. A closer inspection shows, that in order to provide boundary conditions for all $x$, one has to be a bit more careful so as not to leave the range of validity of the linear Eq.~(\ref{Pdotsimple}), i.e., the range where $|m|>1-\epsilon$ for some fixed $\epsilon$ (we usually imposed at least $\epsilon<10^{-4}$). The latter requires a stepwise forward-integration from $x_0=0$ to $x_1$, from $x_1$ to $x_2$ etc., where in each interval $[x_n,x_{n+1}]$ the boundary conditions are imposed at a different (decreasing) $y_n$.

It is usually sufficient to store $P(x_n,|y|<2 y_n)$ both to calculate boundary conditions for $P(x_n<x<x_{n+1},y_n)$ and new initial conditions $P(x_{n+1},|y|<2 y_{n+1})$.
The partitioning $\{x_n\}$ is determined dynamically in such a way as to ensure that $|m(x,y_n-\Delta_n)|>1-\epsilon$ for all $x_n<x<x_{n+1}$ which is required to justify the use of Eq.~(\ref{Pdotsimple}). Here, $\Delta_n=\beta\int^{x_{n+1}}_{x_n} d u\, u \dot{\Lambda}(u)$

With this technique, the whole interval $0<x<1$ can be covered maintaining a high precision in the regime of interest $f(\beta x) y=O(1)$ for the purpose of calculating the overlap $Q(x)$ and to impose self-consistency.
We could reach as low temperatures as $\beta =100$ without encountering numerical problems. In this way we have achieved a high precision for the universal tails in $\Lambda(b), Q(b), r_{\rm sc}(b)$ for $b\gg 1$.


\begin{thebibliography}{68}
\bibitem{fn1}
{In the electron problem, the total number of electrons and thus the total pseudospin is conserved in the canonical ensemble which imposes a certain constraint on the microscopic dynamics. We believe that for long time scales corresponding to inter-state transitions (during which single active electrons can travel much further than a correlation length), this difference with standard Langevin dynamics of spins plays a role. It is these time scales that are encoded in the non-trivial replica structure, as reviewed in Sec.~\ref{ss:InterpretationofRSB}.}
\bibitem{fn3}
{In Ref.~\onlinecite{MuellerIoffe04} the cubic terms  ${\rm Tr} (\delta \Lambda)^3$ and $\sum_{ab} (\delta\Lambda_{ab})^3$
were omitted. While the first term cannot change the nature of the transition locally, the latter could in principle drive the transition discontinuous, as it happens, e.g., in the Pott's glass with $q\geq 4$ colors. However, we checked that this is not the case in the models under study here, neither for the 3D electron glass nor for the random field SK model.}
\bibitem{fn4}
{At infinitely long times the overlap of the actual configuration with the initial state will thus asymptotically decay to the smallest possible overlap of states, $Q(0)$, instead of approaching the equilibrium overlap $\int dx Q(x)$. The latter would be expected if one started from an equilibrated state}.
\bibitem{fn5}
{Even though our numerical iteration starts with a continuous parametrization of $\Lambda(x)$ it does not exclude a self-consistent solution with steps. Such a solution would be signaled in the iteration process by smooth steps which would sharpen under iteration.}
\end{thebibliography}

\end{document}